\documentstyle[12pt,epsfig]{article}

\newcommand{\be}{\begin{eqnarray}}
\newcommand{\ee}{\end{eqnarray}}

\def\anue{{\bar\nu_e}}

\newcommand{\dm}{\mbox{$\Delta m_{21}^2$~}}

\newcommand{\br}{\mbox{$^{8}{B}~$}}

\newcommand{\kl}{\mbox{KamLAND~}}

\newcommand{\thsol}{\mbox{$\theta_{12}$~}}

\newcommand{\ms}{\Delta m^2_{21}}

\newcommand{\sss}{\sin^2 \theta_{12}}
\newcommand{\sch}{\sin^2 \theta_{13}}
\def\ltap{\ \raisebox{-.4ex}{\rlap{$\sim$}} \raisebox{.4ex}{$<$}\ }
\def\gtap{\ \raisebox{-.4ex}{\rlap{$\sim$}} \raisebox{.4ex}{$>$}\ }

\begin{document}

\begin{flushright}
SISSA 77/2004/EP     \\
hep-ph/0410283
\end{flushright}

\begin{center}
{\Large \bf  
High Precision Measurements of $\theta_{\odot}$ in  
Solar and Reactor Neutrino Experiments
}
\vspace{.5in}

{\bf Abhijit Bandyopadhyay$^1$,
Sandhya Choubey$^{2,3}$,
Srubabati Goswami$^4$,
S.T. Petcov$^{3,2,5}$
}
\vskip .5cm

$^1${\it Theory Group, Saha Institute of Nuclear Physics,
1/AF, Bidhannagar,
Calcutta 700 064, India}\\
$^2${\it INFN, Sezione di Trieste, Trieste, Italy.}\\
$^3${\it Scuola Internazionale Superiore di Studi Avanzati, 
I-34014 Trieste, Italy.}\\
$^4${\it Harish-Chandra Research Institute, Chhatnag Road, Jhusi,
Allahabad  211 019, India.}\\
$^5${\it Institute of Nuclear Research and
Nuclear Energy, Bulgarian Academy of Sciences, 1784 Sofia, Bulgaria.}

\vskip 1in

\end{center}

\begin{abstract}
We discuss the possibilities of high precision
measurement of the solar neutrino 
mixing angle $\theta_\odot \equiv
\theta_{12}$ in solar and reactor 
neutrino experiments. The improvements in 
the determination of $\sin^2\theta_{12}$,
which can be achieved with the expected increase
of statistics and reduction of systematic
errors in the currently operating 
solar and KamLAND experiments, are summarised.
The potential of LowNu $\nu-e$ elastic
scattering experiment, designed to measure the $pp$ solar 
neutrino flux, for high precision 
determination of $\sin^2\theta_{12}$, 
is investigated in detail. The accuracy 
in the measurement of $\sin^2\theta_{12}$, 
which can be achieved in a reactor experiment 
with a baseline $L \sim (50-70)$ km, corresponding 
to a Survival Probability MINimum (SPMIN), 
is thoroughly studied. We include the effect of 
the uncertainty in the value of $\sin^2\theta_{13}$
in the analyses. A LowNu measurement
of the $pp$ neutrino flux with a 1\% error
would allow to determine $\sin^2\theta_{12}$
with an error of 14\% (17\%) at 3$\sigma$ 
from a two-generation (three-generation) analysis. 
The same parameter $\sin^2\theta_{12}$ can be measured with
an uncertainty of  2\% (6\%) at 1$\sigma$ (3$\sigma$)
in a reactor experiment with $L \sim60 $ km, 
statistics of $\sim$60 GWkTy and 
systematic error of 2\%.
For the same statistics, the increase of 
the systematic error from 2\% to 5\% leads to an increase
in the uncertainty in  $\sin^2\theta_{12}$
from 6\% to 9\% at 3$\sigma$.
The inclusion of the 
$\sin^2\theta_{13}$ uncertainty in the analysis 
changes the error on $\sin^2\theta_{12}$ to 3\%  (9\%).
The effect of $\sin^2\theta_{13}$ uncertainty
on the $\sin^2\theta_{12}$ measurement 
in both types of experiments
is considerably smaller than naively expected.
\end{abstract}

\newpage

\section{Introduction}
 There has been a remarkable progress in the studies of neutrino
oscillations in the last several years.
The experiments with solar, 
atmospheric and reactor neutrinos 
\cite{cl,ga,sksolar,Ahmad:2002jz,Ahmed:2003kj,
kearns04,kl162} 
have provided compelling evidences for the 
existence of neutrino oscillations 
driven by non--zero neutrino masses and neutrino mixing.
Evidences for oscillations of neutrinos were
obtained also in the first long baseline
accelerator neutrino experiment K2K \cite{K2K}.

The recent Super-Kamiokande  data on the 
$L/E$-dependence of 
multi-GeV $\mu$-like  atmospheric 
neutrino events \cite{kearns04},
$L$ and $E$ being the distance traveled 
by neutrinos and the neutrino energy,
and the new more precise 
spectrum data of KamLAND and K2K experiments 
\cite{kl766,K2Knu04},
are the latest significant 
contributions to 
this progress.
For the first time the data
exhibit directly the effects of the 
oscillatory dependence on $L/E$ and $E$ of 
the probabilities of  
$\nu$-oscillations in vacuum \cite{BP69}.
We begin to ``see'' the 
oscillations of neutrinos.
As a result of these
magnificent developments, 
the oscillations of solar $\nu_e$,
atmospheric $\nu_{\mu}$ and
$\bar{\nu}_{\mu}$, 
accelerator $\nu_{\mu}$ (at $L\sim$250 km)
and reactor $\bar{\nu}_e$ (at $L\sim$180 km), 
driven by nonzero $\nu$-masses 
and $\nu$-mixing, can be considered as 
practically established.

  The SK atmospheric neutrino 
and K2K data are best described in 
terms of dominant 2-neutrino
$\nu_{\mu} \rightarrow \nu_{\tau}$ 
($\bar{\nu}_{\mu} \rightarrow \bar{\nu}_{\tau}$)
vacuum oscillations.
The best fit values and the 
99.73\% C.L. allowed ranges of the 
atmospheric neutrino 
oscillation parameters
$|\Delta m^2_{\rm A}| = |\Delta m^2_{31}|$
and $\theta_{\rm A} \equiv \theta_{23}$
read \cite{kearns04}:
$|\Delta m^2_{31}| = 2.1\times10^{-3}~{\rm eV^2}$,~
$\sin^22\theta_{23} = 1.0$,  
$|\Delta m^2_{31}| = (1.3 - 4.2)\times 10^{-3}~{\rm eV^2}$,
$\sin^22\theta_{23} \geq 0.85$.
The sign of $\Delta m^2_{31}$ 
and of $\cos2\theta_{23}$, if
$\sin^22\theta_{23} \neq 1.0$, 
cannot be determined using
the existing data. 

The combined 2-neutrino oscillation 
analysis of the solar neutrino 
and the new \kl 766.3 Ty spectrum  
data shows \cite{kl766,kl766us,kl766others} that the 
solar neutrino oscillation parameters 
lie in the low-LMA region :
$\Delta m^2_{\odot} \equiv \Delta m^2_{21} = 
(7.9^{+0.6}_{-0.5})\times 10^{-5}~{\rm eV^2}$, 
$\tan^2 \theta_\odot \equiv \tan^2 \theta_{12} = 
(0.40^{+0.09}_{-0.07})$ \cite{kl766}.
The high-LMA solution is excluded at 
more than 3$\sigma$. 
The value of $\Delta m^2_{21}$ is 
determined with a remarkably 
high precision of 12\% at 3$\sigma$.
Maximal solar neutrino mixing
is ruled out at $\sim 6\sigma$.

The solar and atmospheric neutrino, and \kl and K2K
neutrino oscillation data require the existence of 
three-neutrino mixing in the weak charged lepton current.
In this case the neutrino mixing is characterised by one 
additional mixing angle $\theta_{13}$ -- the only small mixing 
angle in the PMNS matrix. Three-neutrino 
oscillation analyses 
of the solar, atmospheric and reactor neutrino data 
show that $\sch < 0.05$ \cite{kl766us,kl766others} \footnote {{After 
the new background data published by the \kl  collaboration   
\cite{kl766} the bound changes to $\sch < 0.055$ \cite{kl766us} 
whereas inclusion of the recently published 391 days SNO salt phase data 
\cite{sno05}
gives $\sch < 0.044$ \cite{progress}}}. 
 
 Understanding the origin of the 
patterns of solar and atmospheric neutrino 
mixing and of $\Delta m^2_{21}$
and $\Delta m^2_{31}$, suggested by the data, 
is one of the central problems 
in neutrino physics today. 
A pre-requisite for any progress in our understanding of neutrino mixing 
is the knowledge of the precise values of solar and atmospheric 
neutrino oscillation parameters,
$\theta_{12}$, $\Delta m^2_{21}$,  
and $\theta_{23}$, $\Delta m^2_{31}$ and of $\theta_{13}$.
In the present article we discuss the 
possibilities of high precision
measurement of the solar neutrino 
mixing angle $\theta_{12}$ in  
solar and reactor neutrino experiments. 

The solar neutrino mixing parameter
$\sin^2\theta_{12}$ is  determined 
by the current KamLAND and solar neutrino data 
with a relatively large uncertainty of 
24\% at 3$\sigma$. In the future, more precise
spectrum data from the KamLAND experiment can lead 
to even more accurate determination of the value of
$\Delta m^2_{21}$. However, 
these data will not provide
a considerably more precise
measurement of $\sin^2\theta_{12}$
owing to the fact that the baseline of the
\kl experiment effectively corresponds to a 
$\bar{\nu}_e$ Survival Probability 
MAXimum (SPMAX) \cite{th12,shika2}.
The analysis of the global solar neutrino data 
taking into account a possible reduction of the 
errors in the data from the phase-III of the 
SNO experiment
shows that the uncertainty in the value 
of $\sss$ would still remain well above 15\% 
at $3\sigma$ \cite{skgd}.

  We begin by summarising the results on $\sin^2\theta_{12}$, 
obtained using the current global solar 
and reactor neutrino oscillation data (section 2). 
We consider the improvements in 
the determination of $\sin^2\theta_{12}$,
which can be achieved with the expected increase
of statistics and reduction of systematic
errors in the experiments which are currently operating.
In particular, the effect of \kl data, 
corresponding to a statistics
of 3 kTy, as well as of the data of 
phase-III of SNO experiment, are analysed.

We turn next to future experiments.
We discuss first (in section 3) the possibility of high 
precision determination of $\sin^2\theta_{12}$ in
a LowNu solar neutrino experiment,
designed to measure
the $\sim$MeV and sub-MeV components of the solar neutrino flux: 
$pp$, $pep$, $CNO$, $^7{Be}$.
It is usually suggested that the LowNu 
experiments can provide one of the
most precise measurements of 
the solar neutrino mixing angle \cite{LENS,XMASS,lownu}.
Detailed analysis was carried out in \cite{roadmap}
and it was concluded that a future $pp$ experiment should have accuracy
better than 3\% in order to improve on the knowledge of $\tan^2\theta_{12}$.

We consider a generic $\nu-e$
scattering experiment measuring the $pp$ neutrino flux
and perform a detailed quantitative 
analysis of the precision with
which $\sin^2\theta_{12}$ can be determined in
such an experiment.
We examine the effect of including 
different 
representative
values of the $pp$ neutrino induced 
event rate in 
the $\chi^2$ analysis of the global solar neutrino data. 
Three values (0.68, 0.72, 0.77) of the (normalized) 
event rate from the currently allowed 
3$\sigma$ range are considered.
The error in the measured rate is varied from 1\% to 5\%.
We investigate how much the accuracy 
on $\sin^2\theta_{12}$ improves 
by adding the $pp$ flux data in the analysis.
The dependence of the sensitivity 
to $\sss$ on the central value of the
measured $pp$ flux as well as on the measurement errors is studied.
We compare the precision in $\sss$ expected with assumed data
on $pp$ neutrinos included in the analysis, with the 
sensitivity that can be achieved using prospective results 
from phase-III of SNO experiment and 3 kTy KamLAND data,
and comment on the minimum error required in the LowNu
$pp$ experiments 
to improve the precision of $\sss$ measurement.  
The impact of the uncertainty due to $\theta_{13}$ 
on the allowed ranges of $\sin^2\theta_{12}$ is studied as well.   

In the section 4
we analyse in detail the possibility of a 
high precision determination of
$\sin^2\theta_{12}$ in a reactor experiment
with a baseline of
$L \sim (50 - 70)$ km, 
corresponding to a $\bar{\nu}_e$ 
Survival Probability MINimum (SPMIN). 
That such an experiment 
can provide the highest
precision in the measurement of $\sss$ was
pointed out first in \cite{th12}. 
A rather detailed study of the precision
in $\sss$, which might be achieved 
in an SPMIN experiment with 
the flux of $\bar{\nu}_e$ from the Kashiwazaki 
reactor complex in Japan and $L = 54$ km, 
was performed recently in \cite{minat12}
\footnote{In ref. \cite{minat12} 
this experiment is called SADO.}.
We consider a generic SPMIN reactor  experiment 
with a KamLAND-type detector.
We investigate the dependence of the
precision on $\sin^2\theta_{12}$ which can 
be achieved in such an experiment
on the baseline, statistics 
and systematic errors. More specifically, the
spectrum data is simulated for four different 
true values of $\ms$ and for each of these values 
the optimal baseline at which the most precise
measurement of $\sss$ could be performed 
is determined. We show, in particular, 
that an independent determination of $\ms$ 
with sufficiently high accuracy would allow
$\sss$ to be measured with the highest precision 
over a relatively wide range of baselines. 
The effect of $\sch$ uncertainty on the $\sss$ determination 
is investigated in detail.

The results of the present study are summarised in section 5.
\section{Measuring $\sin^2\theta_{12}$ in Existing Experiments}

In this section we review the precision of $\dm$ and $\sss$ 
determination from the existing solar neutrino and \kl  data 
from two-neutrino oscillation analysis.  
We also discuss possible improvements in the 
precision that can be achieved
in the currently running experiments.
\subsection{Current Solar and Reactor Neutrino Data on $\sin^2\theta_{12}$}

   In the present global solar neutrino and \kl data analysis 
we include 
\begin{itemize}
\item
data from the radiochemical experiments,
Cl \cite{cl} and Ga (Gallex, SAGE and GNO combined) \cite{ga},
\item
the 1496 day 44 bin Zenith
angle spectrum data from SK \cite{sksolar}, 
\item
the  34 bin combined CC, NC and Electron Scattering (ES)
energy spectrum data from the
phase I ( pure $D_2O$ phase) of SNO \cite{Ahmad:2002jz},
\item
the data on CC, NC and ES total observed rates
from the phase II (salt phase) of SNO experiment 
\cite{Ahmed:2003kj}. 
\end{itemize}
The $^8B$ flux normalization factor
$f_B$ is left to vary freely in the analysis,
while for the
$pp$, $pep$, $^7Be$, $CNO$, and $hep$ fluxes the 
predictions and uncertainties
from the recent standard solar model (SSM) 
\cite{bp04} (BP04) are used. 
We skip the details of the $\chi^2-$ 
analysis, 
which can be 
 found 
in \cite{snocc,snonc,snosaltus}.  

  We include the 766.3 Ty \kl data  in the global analysis. 
For treatment of the latest \kl data in the 
combined analysis we refer the reader to
\cite{kl766us}, while details regarding 
the future-projected analysis with 
simulated \kl data are given in \cite{prekl}.
The best-fit in the combined analysis of solar and 
\kl data is obtained for \cite{kl766us} 
\footnote{With the inclusion of the new 
\kl background data \cite{kl766},
the best-fit shifts to slightly smaller values of 
$\Delta m^2_{21}$:   
$\Delta m^2_{21} = 8.0 \times 10^{-5}$ eV$^2$,~~~ $\sin^2\theta_{12}=0.28$,
~~$\mathrm f_B$ = 0.88 \cite{kl766us}. The inclusion of the recent SNO results 
give the best-fit parameters as 
$\Delta m^2_{21} = 8.04 \times 10^{-5}$ eV$^2$,~~~ $\sin^2\theta_{12}=0.31$,
\cite{progress}
}
\begin{itemize}
\item
$\Delta m^2_{21} = 8.4 \times 10^{-5}$ eV$^2$,~~~ $\sin^2\theta_{12}=0.28$,
~~~$\mathrm f_B$ = 0.88
\end{itemize}
\begin{figure}[t]
\begin{center}
\includegraphics[width=14.0cm, height=10cm]{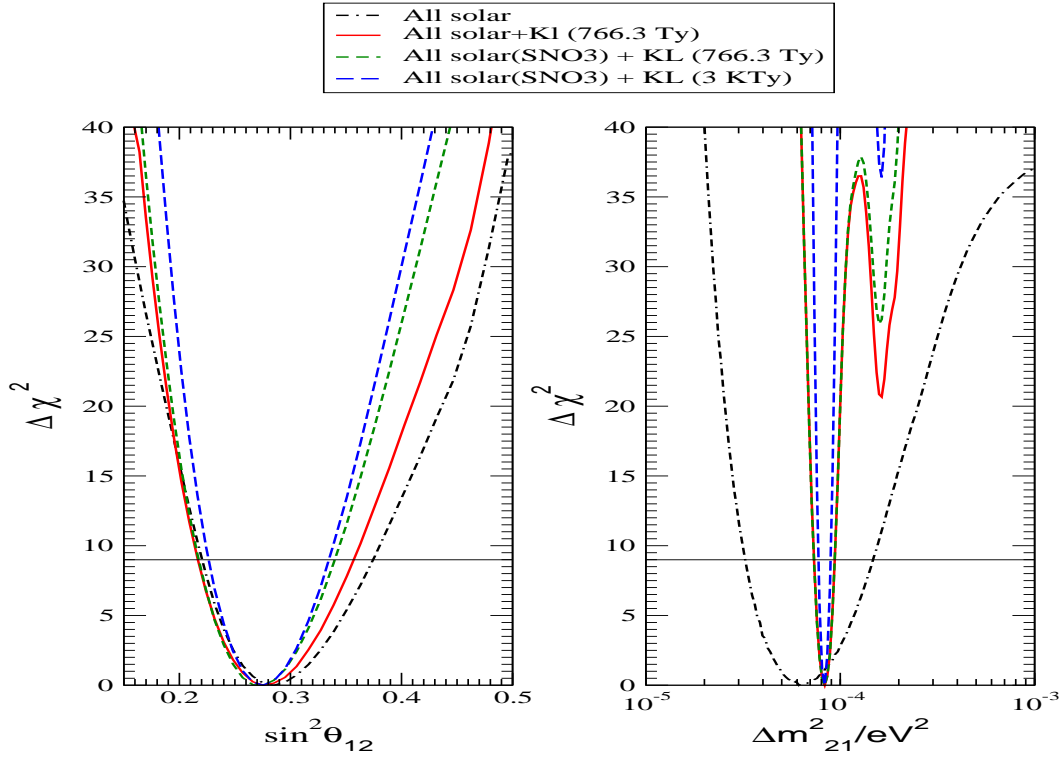}
\caption{$\Delta \chi^2$ as a function of $\sin^2\theta_{12}$ 
(left panel) and of $\dm$ (right panel). 
The curves shown are obtained from global analysis of
the current solar neutrino data (black dot-dashed line), 
current solar + \kl data (red solid line), 
solar neutrino data with projected SNO-III errors + current \kl data 
(green short dashed line) and the 
solar neutrino data with projected SNO-III errors + 
prospective 3 kTy \kl data (blue long dashed line).
The horizontal lines indicate the 3$\sigma$ limit ($\Delta \chi^2 = 9$)
for 1 parameter fit.}
\label{delchi}
\end{center}
\end{figure}

In Fig. \ref{delchi} we plot $\Delta \chi^2 =  \chi^2 - \chi^2_{min}$ 
as a function of $\dm$ (right panel) and $\sss$ (left panel),
for a two-neutrino oscillation 
fit of the global solar neutrino + \kl data.  
The parameters not given in each of the two
panels  are allowed to vary freely.  
From this figure one can easily read off the  allowed ranges of the 
displayed parameter at various confidence levels. 
The horizontal line shows the 3$\sigma$ limit corresponding to 
$\Delta \chi^2 =9$ for a 1 parameter fit. 
In Table 1 we present the current 3$\sigma$ allowed   
ranges of $\dm$ and $\sin^2\theta_{12}$ obtained from 
Fig. \ref{delchi}.
We also give the spread in these parameters which is defined as 
\be
{\rm spread} = \frac{ prm_{max} - prm_{min}}
{prm_{max} + prm_{min}}\times 100
\label{error}
\ee
where ${prm}$ is either \dm or $\sin^2\theta_{12}$.    
Table 1 and Fig. 1 demonstrate clearly that
the $\kl$ experiment has remarkable sensitivity to $\dm$. 
The inclusion of \kl data with 
increasing statistics in the analysis 
progressively reduces the spread in $\dm$, 
and with the latest data the  3$\sigma$ 
spread is $\sim$ 12\%.
This demonstrates, in particular,
the extraordinary precision 
that has already been achieved in 
the determination of $\ms$. 
On the other hand, the spread in  $\sin^2\theta_{12}$ 
is seen to be controlled mainly by the solar neutrino data and 
does not show any marked reduction with the inclusion of 
the \kl data in the analysis. 
Using the global solar neutrino and \kl 
data allows to determine $\sss$ with an
error of 24\% at 3$\sigma$.
\subsection{Prospective \kl and SNO Data and 
$\sin^2\theta_{12}$}

We will analyze next the expected impact of future data 
from \kl and SNO experiments on the
$\ms$ and $\sss$ determination.

  The SNO experiment is sensitive to the 
flux of $^8B$ neutrinos with energy 
$E\gtap 5$ MeV. For the 
oscillation parameters in the low-LMA region,
the  $^8B$ neutrino survival probability 
of interest is given approximately by
(MSW adiabatic transition probability):  
\be
P_{ee}({\br})\approx \sin^2\thsol.
\label{pee8b} 
\ee 
The CC/NC ratio measured in SNO
determines $P_{ee}$ independently 
of the $^8B$ neutrino flux normalization.
Thus, it can give a direct measure of 
$\sin^2\theta_{12}$.
Reducing the errors ($\Delta({R_{CC}/R_{NC}}$) in 
the measured CC and NC event rates in SNO,
$R_{CC}$ and $R_{NC}$,
can improve the precision of 
determination of this parameter since
\be
\Delta({\sin^2\theta_{12}}) = \Delta({R_{CC}/R_{NC}}). 
\label{errsno} 
\ee

   The oscillations of reactor $\bar{\nu}_e$ 
detected in \kl experiment
are practically not affected by Earth
matter effects and the corresponding $\bar{\nu}_e$ 
survival probability has the form 
\be
P_{\bar{e}\bar{e}}^{KL}\approx1- \sin^22\theta_{12}
\sin^2\left(\frac{\dm L}{4 E}\right).
\label{eq:klpee}
\ee
The average energy and baseline for \kl 
correspond to 
$\sin^2(\dm L/4E) \approx  0$, i.e.,  
to a Survival Probability MAXimum (SPMAX).
As a consequence, the coefficient
of the $\sin^22\theta_{12}$ term
in $P_{\bar{e}{e}}^{KL}$ is relatively small,
weakening the  sensitivity of  \kl  to
$\theta_{12}$. As was shown in \cite{th12},
the most precise measurement of $\sss$  can be 
performed in a reactor experiment with a baseline tuned to 
a Survival Probability MINimum
(SPMIN), i.e., to $\sin^2(\dm L/4E) \approx  1$.
We will discuss the sensitivity to 
$\sin^2\theta_{12}$, which can be achieved in 
such an experiment, in section 4. 

  In phase-III of the SNO experiment, 
the NC events will be observed directly
(and independently from the CC events)
using $^3He$ proportional counters.
This will help to increase the NC statistics and 
reduce the systematic errors in the NC data.
In addition, the correlations between the errors
in the measured CC and NC event rates will be absent.
The total projected error in the
measured NC event rate in phase-III of SNO experiment is 
$\sim 6\%$ \cite{sno3}.
We incorporate this in our 
analysis instead of the present error
in $R_{NC}$ of 9\%. For the CC event rate
$R_{CC}$ measured at SNO
we assume a somewhat reduced total 
error of 5\%  ({{the current error in $R_{CC}$ is 
approximately 6\% \cite{sno05}}}).
We assume also that the central values of the 
measured CC and NC even rates
will remain unchanged.
The results of our analysis are shown in Fig. \ref{delchi}. 
The short dashed lines in Fig. \ref{delchi} 
display the behavior of $\Delta \chi^2$ 
with anticipated
SNO phase-III results and   
the reduced projected errors
added to the global solar neutrino and
\kl 766.3 Ty spectrum data.  
The figure shows that 
with the inclusion of the SNO phase-III 
prospective results, 
the allowed range of $\dm$ 
in the low-LMA region remains unchanged:
it is determined principally by the \kl data.
However, the higher
$\dm$ regions get more disfavored
as the reduced errors 
in $R_{CC}$ and $R_{NC}$
lead to a stronger rejection of 
larger values of $R_{CC}/R_{NC}$ \cite{MMSP02}.
The figure also shows that 
the allowed range of 
$\sin^2\theta_{12}$ gets 
further constrained from above.
From Table \ref{tab_spread} we see 
that the  3$\sigma$ spread 
of $\sin^2\theta_{12}$ becomes 21\% with the 
inclusion of projected SNO phase-III data 
in the presently existing set of data. 

  We also studied 
the effect of increased statistics of \kl
experiment on the $\sss$ and $\ms$ 
determination. To this end, we include 
3 kTy \kl spectrum data, 
simulated at $\dm = 8.3\times 10^{-5} $ eV$^2$,
in our analysis with 
projected SNO phase-III data.
We use a systematic error of 
5\% for \kl since 
the \kl systematic error is 
expected to diminish
\footnote{{The \kl systematic 
error at present is 7.13\% including the ($\alpha$-n) background and 
6.5\% without this background  \cite{kl766}.}}
after the planned fiducial volume 
calibration and 
re-evaluation of the uncertainties
in the power of the relevant nuclear stations.
If the real \kl spectrum data 
conforms to this projected spectrum,
the allowed range of $\dm$ would be 
further constrained, allowing the  
determination of $\dm$ with an accuracy of about 5\%.
The higher \dm regions would be disfavored
even stronger. 
The uncertainty in the value of $\sin^2\theta_{12}$ 
would be smaller
and the 3$\sigma$ spread, as seen from 
Table \ref{tab_spread}, could be 18\%. 
Clearly, the uncertainty in $\sin^2\theta_{12}$ 
cannot be reduced to 15\% or less (at 3$\sigma$)
by future data from the currently operating 
solar and reactor neutrino experiments 
\footnote{{If we include the recently published SNO results \cite 
{sno05} in our analysis then 
the 3$\sigma$ spread in $\sss$ becomes 26\%  whereas the solar+\kl analysis 
including the recent SNO results and the $(\alpha,n)$ background in \kl 
\cite{kl766} gives the spread as 22\% \cite{progress}.}}

\begin{table*}[htb]
\begin{tabular}{ccccc}
\hline
{Data set} & (3$\sigma$)Range of & (3$\sigma$)spread in  &
(3$\sigma$) Range of
&(3$\sigma$) spread in \cr
{used} & $\Delta m^2_{21}$ eV$^2$
& {$\Delta m^2_{21}$} & $\sin^2\theta_{12}$
& {$\sin^2\theta_{12}$} \cr
\hline
{only sol} & 3.3 - 15.3 &{65\%} & $0.22-0.38$ &27\%\cr
{sol+ 766.3 Ty KL}& 7.4 - 9.5 & 12\% & $0.22-0.36$ & 24\% \cr
{sol(+SNO3) + 766.3 Ty KL} & 7.4 -9.5 & 12\%  & 0.22 - 0.34 & 21\% \cr
{sol(+SNO3)+3KTy KL}  & 7.7 -8.9 & 7\%& 0.23 - 0.33 & 18\% \cr \hline
\end{tabular} \\[2pt]
\caption{
The 3$\sigma$ allowed ranges and spread (in per cent)  
of $\Delta m^2_{21}$ and
$\sin^2\theta_{12}$ obtained from  1 parameter fits.
}
\label{spread}
\label{tab_spread}
\end{table*}
\section{Determining $\theta_{12}$ from Measurement of 
the $pp$ Solar Neutrino Flux}

 The $pp$ fusion reaction is the main contributor to 
the observed solar luminosity and the corresponding $pp$ 
neutrinos constitute the 
largest and dominant component of the solar neutrino flux. 
The SSM uncertainty in the predicted solar neutrino fluxes is the  
least for the $pp$ neutrinos ($\sim$1\%) \cite{bp04}. 
So far only the Ga experiments have 
provided 
information on the $pp$ neutrino 
flux because of their 
low energy threshold of 0.23 MeV. 
Sub-MeV solar neutrino experiments 
(LowNu experiments) are being planned
for measuring the flux of $pp$ neutrinos
using either charged
current reactions (LENS, MOON, SIREN) or
$\nu - e^-$ elastic scattering
(XMASS, CLEAN, HERON, MUNU, GENIUS)
\cite{LENS,XMASS,lownu}.  
Since according to the SSM, 
most of the energy released by the Sun 
($\sim$99\%) is 
generated in the $pp-$cycle of 
reactions in which also
the $pp$ neutrinos are produced, 
a precise determination of the $pp$ neutrino flux 
would lead to a better understanding of the 
solar energetics and, more generally,
of the physics of the Sun.
It has also been realized that 
high precision measurement of the 
$pp$ neutrino flux can be instrumental for  
more accurate determination of 
the solar neutrino mixing parameter, 
which, as we have seen in the preceding section, 
will not be determined 
with an uncertainty smaller than
$\sim$18\%
(at 3$\sigma$) by the 
currently  operating experiments.  

    Since the $pp$ neutrino energy 
spectrum extends upto 0.42 MeV only, for $\ms$ 
in the
LMA region, the $pp$ neutrino
oscillations are practically not affected
by matter effects 
in the Sun or 
the Earth.
Thus, to a good approximation, the
$pp$ neutrino oscillations 
are described by the
$\nu_e$ survival 
probability in the case of
oscillations in vacuum,
in which the oscillating term  
is strongly suppressed by the averaging 
over the region of neutrino
production in the Sun \cite{SPJRich89}:
\be
P_{ee}^{2\nu}(pp) \cong 1 - \frac{1}{2}\sin^22\theta_{12}.
\label{peepp}
\ee
The normalised event rate for a $\nu-e$ scattering (ES)
and charged current (CC) experiments measuring the 
$pp$ neutrino flux
is given respectively by
\be
R_{pp} &=& \langle{P_{ee}^{2\nu}(pp)}\rangle + r_{pp} 
(1 - \langle{P_{ee}^{2\nu}(pp)}\rangle)
~~{\mathrm{ for~~ ES~~ experiments}},
\label{rppes}
\\
R_{pp} &=& \langle{P_{ee}^{2\nu}(pp)}\rangle 
~~~~~~~~~~~~~~~~~~~~~~~~{\mathrm{ for~~ CC~~ experiments}},
,
\label{rppcc}
\ee
{ where $r_{pp} = \int f_{pp}(E)\sigma(\nu_{\mu(\tau)}e^-)dE/
\int f_{pp}(E) \sigma(\nu_{e}e^-)dE 
\approx 0.3$ 
where $f_{pp}(E)$ is the pp-neutrino flux,
$\sigma(\nu_{\mu(\tau)}e^-)$ and $\sigma(\nu_{e}e^-)$
are the $\nu_{\mu(\tau)} - e^-$ and $\nu_{e} - e^-$
eleastic scattering cross sections
and $\langle..\rangle$ denotes averaged probabilites.
\footnote{{
Strictly speaking, eq. (6) is an approximate expression since
$P^{2\nu}_{ee}(pp)$ depends on the neutrino energy, while eq. (6)
is valid for energy-independent $P^{2\nu}_{ee}(pp)$.
In our numerical analysis
we have not used this approximation.
.}}
}

For an ES experiment, the second term in eq. (\ref{rppes})
represents the NC contribution. 
Since the pp neutrino survival probability is largely independent of 
energy one can use eq. {\ref{peepp}} for the averaged probabilites .  
Using eqs. (\ref{peepp}) - (\ref{rppcc}), one finds for
the uncertainty in $\sss$ determination: 
\be 
\Delta({\sss})^{CC}_{pp} \sim \frac{\Delta{R_{pp}}} {2 \cos2\theta_{12}}
~~{\mathrm{ for~~ CC~~ experiments}}
\label{errppcc}
\\
\Delta({\sss})^{ES}_{pp} \sim \frac{\Delta{R_{pp}}} {2 \cos2\theta_{12}} 
\frac{1}{1-r_{pp}}
~~{\mathrm{ for~~ ES~~ experiments}}
\label{errppes}
\ee
where $\Delta{R_{pp}}$ is the error in the 
measured value of $R_{pp}$.
A comparison of eq. (\ref{errppcc}) with 
eq. (\ref{errsno}) shows that 
for the same value 
of $\Delta{R_{pp}}$
and of the error in the CC to NC event rate 
ratio measured by SNO,
$\Delta({R_{CC}/R_{NC}})$,
the $pp$ neutrino experiments of the CC type can provide 
a more precise measurement of $\sss$ only  
if $\cos2\theta_{12} > 0.5$ ($\sss < 0.25$). 
Similarly, it follows from 
eq. (\ref{errppes}) and eq. (\ref{errsno}) that
for $\Delta{R_{pp}} \cong \Delta({R_{CC}/R_{NC}})$,
the ES $pp$ experiment could provide a more precise
determination of $\sss$ only 
if the true value of $\cos2\theta_{12} > 0.71$ ($\sss < 0.14$).
Since the currently allowed 3$\sigma$ 
range of $\sss$ is $\sss = (0.22-0.38)$, 
for almost all of the allowed values of $\sss$,
SNO will have a better 
sensitivity to $\sss$ 
than a LowNu $pp$ experiment 
measuring the $pp$ neutrino flux with
the same experimental error as
the error in the SNO data on
the ratio $R_{CC}/R_{NC}$. 

 In order to improve the accuracy of $\sss$ 
determination after the SNO phase-III results 
will be available, the total experimental error 
in the measured event rate in the $pp$ experiments
has to be sufficiently small, which 
requires 
 high statistics and 
well understood systematics. 
In the present  section  we will quantify these
statements by incorporating 
hypothetical data on the $pp$ neutrino flux in our analysis. 
We consider a generic $pp$ neutrino ES experiment
\footnote{A comparison of eqs. (\ref{errppcc}) and (\ref{errppes}) 
shows that the CC $pp$ experiments could achieve a better sensitivity 
on $\sss$ due to the absence of the $\approx 30\%$ 
``contamination'' caused by the NC events 
present in the ES event sample. 
In what follows, we present results for an ES $pp$ 
neutrino experiment (XMASS, etc.).
Henceforth the term ``$pp$'' experiment 
implies an ES $pp$ neutrino experiment.}
and consider some illustrative sample rates from the 
currently predicted range. 
We give quantitative estimate 
of the sensitivity to $\sin^2\theta_{12}$ 
expected to be achieved in 
i) a $pp$ experiment, and
ii) combining the prospective data
from a $pp$ experiment with the 
global solar and reactor
neutrino data.
In particular, we estimate the 
maximal error in a $pp$ flux measurement, 
for which the uncertainty in the 
determined value of $\sin^2\theta_{12}$ 
would be smaller than that expected 
after the inclusion of the SNO phase-III results.
We use in this analysis 
the $pp$ neutrino flux 
and its 1\% uncertainty predicted 
by the BP04 SSM \cite{bp04}.
The 1\% error due to SSM uncertainties
is added to the experimental errors
in the $pp$ flux determination. 
\subsection{Two Generation Analysis} 
\begin{figure}[t]
\includegraphics[width=10.0cm, height=14cm,angle=270]{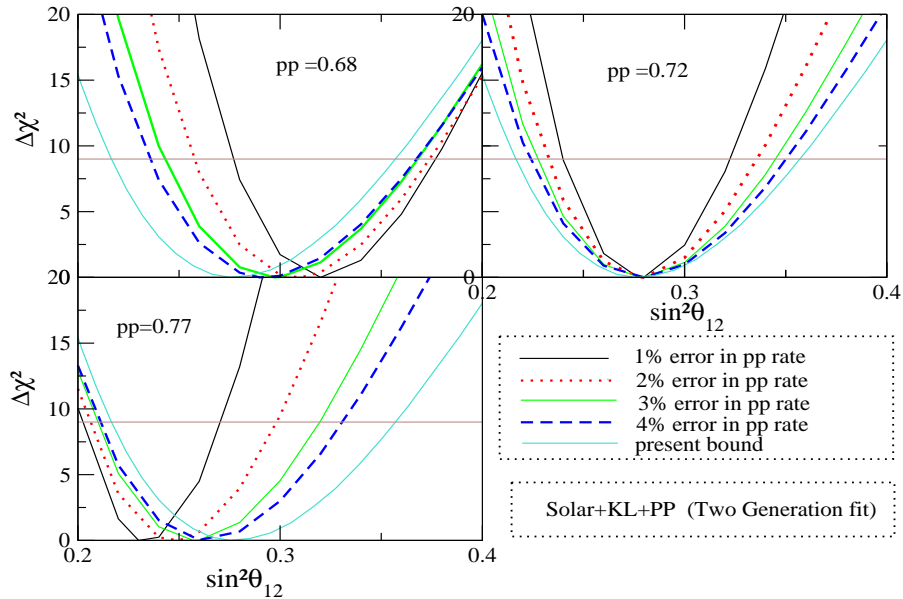}
\caption{$\Delta \chi^2$ as a function of $\sin^2\theta_{12}$. 
The results shown are obtained in a 
two-neutrino oscillation analysis of the
global solar neutrino and 
\kl 766.3 Ty spectrum data and 
simulated data from a 
LowNu $pp$ neutrino experiment. 
The three panels correspond to three 
illustrative values of the 
event rate due to $pp$ neutrinos 
in the LowNu experiment (``$pp$ rate''), 
normalized to the rate
predicted by the BP04 SSM.  
In each case results for 
four different assumed values 
of the  error in the measured $pp$ rate are shown.
We also show the curve obtained in global analysis of
the current solar and reactor neutrino data.
The horizontal line indicates the 3$\sigma$ limit 
($\Delta \chi^2 = 9$)
for 1 parameter fit.}
\label{fig2}
\end{figure}
\begin{figure}[t]
\includegraphics[width=14.0cm, height=10cm]{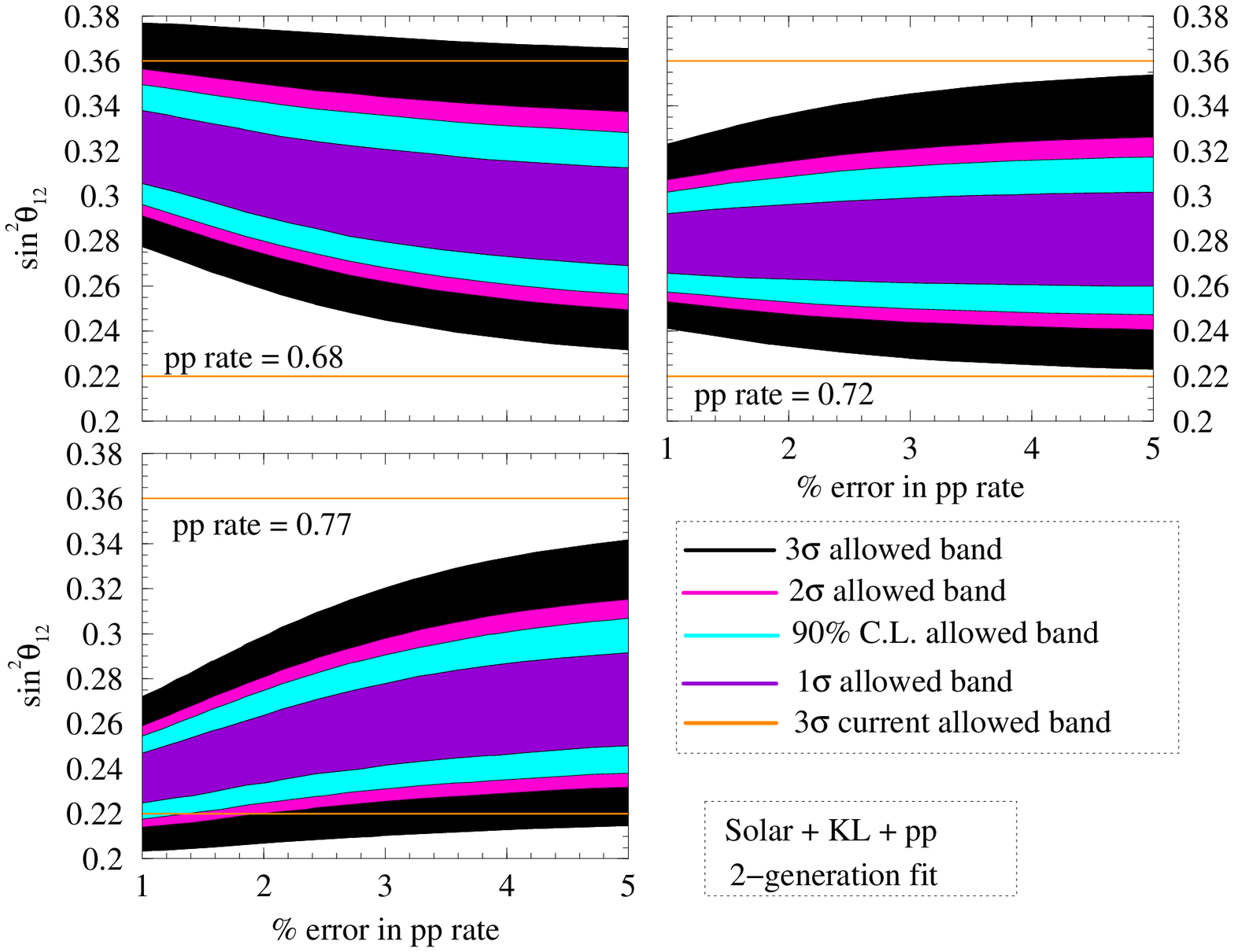}
\caption{The 1$\sigma$ (68.27\% C.L.), 
1.64$\sigma$ (90\% C.L.),
2$\sigma$ (95.45\% C.L.) and  3$\sigma$ (99.73\% C.L.)
allowed range of $\sin^2\theta_{12}$ as a 
continuous function 
of the error in the $pp$ rate
for three assumed mean values of 
the measured $pp$ rate.
Shown is also the current 
3$\sigma$ allowed range of $\sss$.
}
\label{fig3}
\end{figure}

   We consider a generic $\nu -e$ 
scattering experiment which can measure the
$pp$ neutrino flux. The experiment is assumed to
have $e^-$ kinetic energy threshold of 50 keV.
We suppose that the BP04 SSM predicts correctly the $pp$ flux.
The predicted event rate (``$pp$ rate'') 
in such an experiment for the best 
fit values of the oscillation parameters, 
normalized to the rate predicted
by the BP04 model in the absence of 
neutrino oscillations, is 0.71;
the predicted 3$\sigma$ range for the normalized
$pp$ rate is  $0.67-0.76$ \cite{kl766us}.
We will consider three illustrative 
values of the normalized $pp$ rate, 
0.68, 0.72 and 0.77, 
and
vary the 
 experimental error 
in the measured $pp$ rate from 1\% to 5\%.
The estimated theoretical uncertainties 
due to the SSM and 
their correlations 
for the $pp$ neutrino flux
are included in the analysis following the
standard covariance approach \cite{flap}. 
Thus, we minimise the $\chi^2$ defined as
\be
\chi^2_{\odot} = \sum_{i,j=1}^N (R_i^{\rm expt}-R_i^{\rm theory})
(\sigma_{ij}^2)^{-1}(R_j^{\rm expt}-R_j^{\rm theory})
\label{chi2}
\ee
where $R_{i}$ are the solar neutrino 
data points, $N$ is
the number of data points and
$(\sigma_{ij}^2)^{-1}$ 
is the inverse of the covariance matrix,
containing the squares of the 
correlated and uncorrelated experimental
and theoretical errors.
The $^8B$ flux normalisation factor
$f_B$ is left to vary freely in the analysis. The errors and 
correlations due to the 
other fluxes are taken from the BP04 SSM
\footnote{One can also keep the normalization of the $pp$ neutrino flux 
as a variable parameter, subject to the solar luminosity constraint
\cite{roadmap}.}.

 Let us begin by analysing first the potential results  
from a possible future $pp$ neutrino experiment alone. 
In this case  eq. (\ref{rppes}) can be used to get the 
approximate values of $\sin^22\theta_{12}$ for a given $pp$ rate,
\be
\sin^22\theta_{12} \approx \frac{2(1-R_{pp})}{1-r_{pp}}.
\label{s12frompp}
\ee
If we assume a $pp$ rate of 
0.72 and 1\% experimental error 
(in addition to the SSM uncertainty of 1\%), 
the $\chi^2$ analysis gives for 
the $3\sigma$ range of allowed values of $\sss$ 
$0.21 < \sss < 0.33$.
This agrees very well with what 
one would obtain using eq. (\ref{s12frompp})
including the errors along with the mean 
$pp$ rate value of 0.72.
Thus, the spread in $\sss$ is of about 22\%, 
which is not much smaller than the spread 
in $\sss$ determined using
the current data. 
In what follows, we will perform an 
analysis of the data from all experiments, 
including 
illustrative  rates from a $\nu-e$ scattering type $pp$ experiment  
, when estimating the 
sensitivity to the mixing angle $\theta_{12}$. 

 Figures \ref{fig2} and \ref{fig3} show results obtained 
in a two-neutrino oscillation analysis
of the \kl and global solar 
neutrino data, including the $pp$ rate
assumed to be measured in the LowNu ES experiment.
In Fig. \ref{fig2} we plot the dependence of 
$\Delta \chi^2$  on $\sin^2\theta_{12}$
and  show results for four values of the error 
in the measured $pp$ rate, 1\%, 2\%, 3\% and 4\%. 
The cyan colored lines correspond to
results obtained using 
the currently existing data.  
One can easily read from the figure
the range of allowed values of $\sss$ at 
a given C.L. for a chosen 
experimental error in the $pp$ rate measurement.
In Fig. \ref{fig3} the corresponding 
allowed range of $\sss$ 
is shown as a function of 
the error in the $pp$ rate measurement. 
We let the $pp$ rate error vary from 1\% to 5\%.
The various bands correspond to 68.27\% C.L. ($1\sigma$), 
90\% C.L. ($1.64\sigma$),
95.45\% C.L. ($2\sigma$) and 99.73\% C.L. ($3\sigma$). 
For comparison, the  $3\sigma$ range 
of the presently allowed values of $\sss$ is indicated 
in the figure by horizontal lines. 
 
 In both figures we present results for 
the 3 assumed values of the measured $pp$ rate, $R_{pp}$.
The figures show that: 
\begin{enumerate} 
\item 
For $R_{pp} = 0.68$ at
the lower end of the predicted range of values of 
the $pp$ rate, the lower bound on $\sin^2\theta_{12}$ 
increases considerably with reducing the error in 
the $pp$ flux measurement. The upper 
bound also increases but not significantly: 
the reduction of the error in $pp$ rate measurement 
has a much smaller effect on the maximal allowed 
value of $\sss$. 
\item
For $R_{pp}= 0.72$, which is close to 
the best-fit predicted rate, both the lower 
bound on $\sss$ increases and the upper bound
decreases, tightening the allowed range of 
$\sss$. Reducing the error in the 
$pp$ flux measurement improves the precision 
of determination of $\sss$.
\item
For the relatively high $pp$ rate, $R_{pp} = 0.77$, 
the minimal allowed value of $\sss$
diminishes somewhat,
while the maximal allowed value
diminishes considerably.
\end{enumerate}
These features can be understood by analyzing the 
expression for the probability  of survival 
of $pp$ neutrinos 
given in eq. (\ref{probpp}).
It is evident that a lower $pp$ rate drives  
$\sss$ towards higher values and  vice versa.
Therefore the lower bound on $\sss$ 
increases for $R_{pp} = 0.68$.  
The effect of reducing the error in the 
$pp$ flux measurement 
has the effect of pushing $\theta_{12}$ 
towards higher values. 
Likewise, one could expect that the 
maximal value of $\sss$ should 
equally increase for 
$R_{pp} = 0.68$.
However, since we have used
the global solar neutrino data in 
the analysis, the maximal allowed 
value of $\sss$ 
increases only slightly
as the corresponding higher values 
of $\sin^2\theta_{12}$ are strongly 
disfavored by the already existing data. 
Thus, if a future $pp$ (ES) experiment measures 
a value of $R_{pp}$ near the 
lower end of the presently predicted range, 
the lower limit on $\sss$ will increase 
as the error in the measured $R_{pp}$ is reduced.
The upper limit will increase slightly, 
but the effect of reducing the $pp$ 
rate error will not be drastic.  

A higher measured value of $R_{pp}$, 
$R_{pp} = 0.77$, requires a 
lower value of $\sss$.  Consequently, 
the maximal allowed value of $\sss$ is seen to 
diminish substantially. 
The lower limit on $\sss$ could be pushed 
to relatively small values by the 
data from the $pp$ experiment alone.
However, such small values of $\sss$ are already 
excluded by the current set of data  
and therefore the lower 
limit on $\sss$ cannot reduce much.

  A $pp$ rate of $R_{pp} = 0.72$ is quite 
consistent with the current best-fit values of the 
parameters. As a consequence, the corresponding
maximal allowed value of $\sss$ diminishes 
and the minimal value increases as the error in 
$R_{pp}$ is reduced.
Thus, the precision of $\sss$ determination increases.

     We summarize in Table 2 
(columns 3 and 4)
the range 
of allowed values of $\sss$ 
expected from a combined analysis of \kl and
global solar neutrino data and the future 
(hypothetical) data from the $pp$ experiment.
With a 1\% experimental error in the $pp$ rate,
the 3$\sigma$ spread can decrease to about 14\%. 
We note that the maximal and/or minimal allowed values 
of $\sss$ depend critically on the measured mean 
value of the $pp$ rate. 
However, the spread in $\sin^2\theta_{12}$ 
is practically independent of the 
latter. For $R_{pp}=0.72$, the 
$3\sigma$  spread in the value of $\sss$ is  
approximately 23\%, 19\%, 18\% and 14\% respectively 
for an error in the $pp$ rate of 4\%, 3\%, 2\% and 1\%  
(Table 2, 4th column). 
{{
The 3$\sigma$ spread in $\sin^2\theta_{12}$ without 
including the phase-III SNO data
is 24\% (Table 1). 
We observe that 
$pp$ flux measurement  
can improve 
the precision of $\sss$ determination even with an error $>$ 4-5\%. 
}}
\begin{figure}[t]
\includegraphics[width=14.0cm, height=10cm]{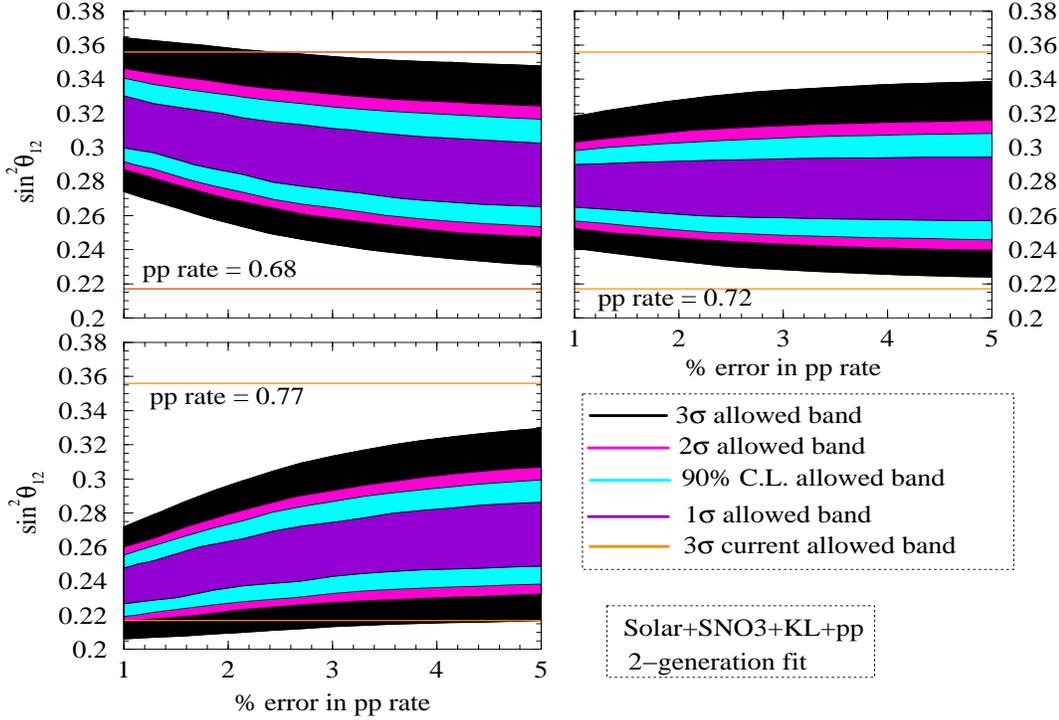}
\caption{
Same as Fig. \ref{fig3} but 
including the prospective data (and errors) from the
phase-III of SNO experiment.
}
\label{fig4}
\end{figure}
In Fig. \ref{fig4} we show the expected range of allowed values
of $\sin^2\theta_{12}$ as a function of the
error in the measured $pp$ rate, after
the potential results and projected errors 
from phase-III of the SNO experiment (SNO3) 
are included in the global analysis. 
If the measured mean $pp$ rate is 0.68 (0.77), the 
maximal (minimal) allowed value of $\sss$ slightly 
increases (decreases) and the minimal (maximal) 
value increases (decreases) significantly. 
The total uncertainty in 
$\sin^2\theta_{12}$ is independent of the central 
value of the $pp$ rate, in agreement, with 
what we have found earlier. For $R_{pp}=0.72$, the 
$3\sigma$ spread in $\sss$ is 
$\approx$ 21\%, 18\%, 18\%, 14\%  
for a 4\%, 3\%, 2\% and 1\% experimental error in $R_{pp}$,
respectively. 

  It follows from our analysis 
that the data from the $pp$ experiment
can allow to reduce the expected uncertainty
of 21\% in the determination of $\sss$ 
after the prospective phase-III SNO results are included in the analysis 
only if the error in the measured $pp$ rate
does not exceed $\sim$ 4\%. 
However, even with $\Delta R_{pp} = 1\%$,
the 3$\sigma$ spread in the value of $\sss$ 
would not be smaller than $\sim$14\%. 

    Table \ref{pp2gspread} summarises 
the results on the allowed ranges and spread 
of $\sin^2\theta_{12}$, obtained in a two-neutrino 
oscillation analysis 
including SNO-III projected errors. 
As expected,
the spread in $\sin^2\theta_{12}$ 
reduces with inclusion of the
phase-III SNO data in the analysis. 
Note that the ranges given in 
Table \ref{pp2gspread} are with the present \kl
766.3 Ty spectrum data. 
If we use future higher statistics data from \kl, 
the allowed spread 
in $\sin^2\theta_{12}$ may reduce somewhat.
 

\begin{table}[htb]
\begin{center}
\begin{tabular}{||c||c||c|c||c|c||} 
\hline \hline
& & \multicolumn{2}{c||}{\rm solar+reactor +pp} & \multicolumn{2}{c||} 
{\rm solar(+SNO3)+reactor+pp} \\ \cline{3-6} 
pp rate &\% error &
3$\sigma$ range & spread & 3$\sigma$ range & spread \\ \hline
0.68 & 1 & 0.28 - 0.38 & 15.2\% & 0.27 - 0.36 & 14.3\%\\
& 2 & 0.26 - 0.37 & 17.5\% & 0.26 - 0.36 & 16.1\% \\
& 3 & 0.24 -0.37 & 21.3\% & 0.24 - 0.35 & 18.6\%  \\
& 4 & 0.24 - 0.37 & 21.3\% & 0.23 - 0.35 & 20.7\%  
\\ \hline
0.72 & 1 & 0.24 - 0.32 & 14.3\% & 0.24 - 0.32 & 14.3\% \\
& 2 & 0.23 - 0.33 & 17.9\% & 0.23 - 0.32 & 16.4\% \\
& 3 & 0.23 - 0.34 & 19.3\% & 0.23 - 0.33 & 17.9\% \\
& 4 & 0.22 - 0.35 & 22.8\%  & 0.22 - 0.34 & 21.4\% 
\\ \hline
0.77 & 1 & 0.20 - 0.27 & 14.9\% & 0.20 - 0.27 & 14.9\% \\
{}   & 2 & 0.21 - 0.30 & 17.6\% & 0.21 - 0.29 & 16.0\% \\
& 3 & 0.21 - 0.32 & 20.8\% & 0.21 - 0.31 & 19.2\% \\
& 4 & 0.21 - 0.33 & 22.2\% & 0.21 - 0.32 & 20.8\% \\
\hline 
\end{tabular}
\end{center}
\caption{
The 3$\sigma$ allowed ranges and \% spread  of 
$\sin^2\theta_{12}$ 
obtained in a 
two-neutrino oscillation analysis of the 
solar neutrino data, 
including the simulated data on the $pp$ neutrino flux,
and the \kl 766.3 Ty spectrum data.
}
\label{pp2gspread}
\end{table}
\subsection{The Impact of Non-Zero $\theta_{13}$}

\begin{figure}[t]
\begin{center} 
\includegraphics[width=14.0cm, height=10cm]{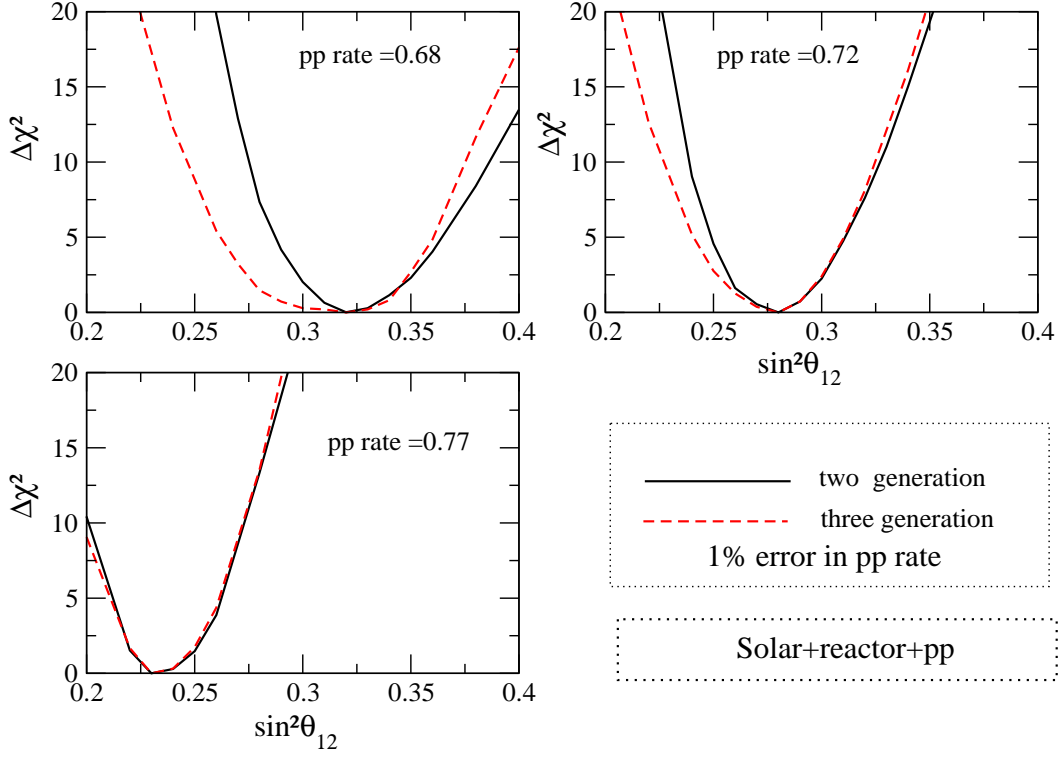}
\caption{\label{delchi2g+3g}
Comparison of $\Delta \chi^2$ vs. $\sss$
obtained from a two generation analysis (solid black line) 
and three generation analysis (dashed red line) analyses
of the world neutrino data including the $pp$ rate.
The three panels show the results for the three 
illustrative values of the $pp$ rate. We present the results 
for 1\% error in the $pp$ rate. For the three-neutrino oscillation
analysis $\sch$ is allowed to vary freely.
}
\end{center}
\end{figure}
\begin{figure}[t] 
\begin{center}   
\includegraphics[width=10.0cm, height=7cm]{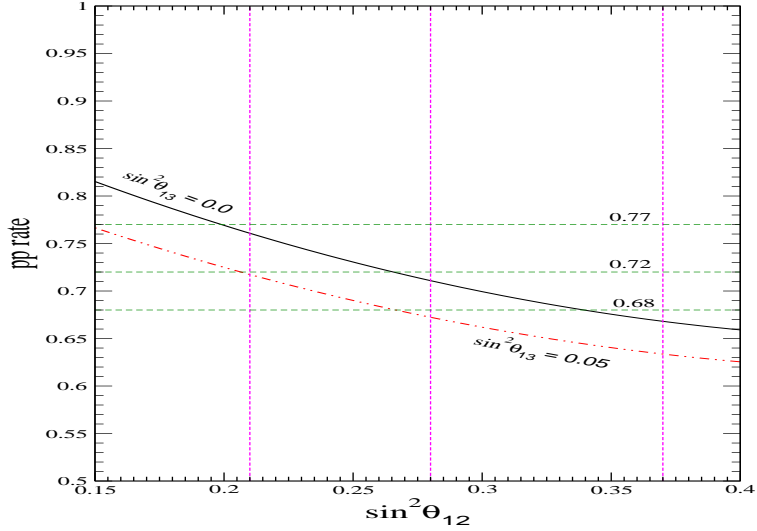}
\caption{\label{probpp}
The $pp$ rate as a function of $\sss$. The black solid line is 
the $pp$ rate for $\sch=0$ while the red dot-dashed line is 
$\sch=0.05$. We also show our illustrative $pp$ rates of 
0.68, 0.72 and 0.77 on the plot.
}
\end{center}
\end{figure}
\begin{figure}[t]
\includegraphics[width=14.0cm, height=10cm]{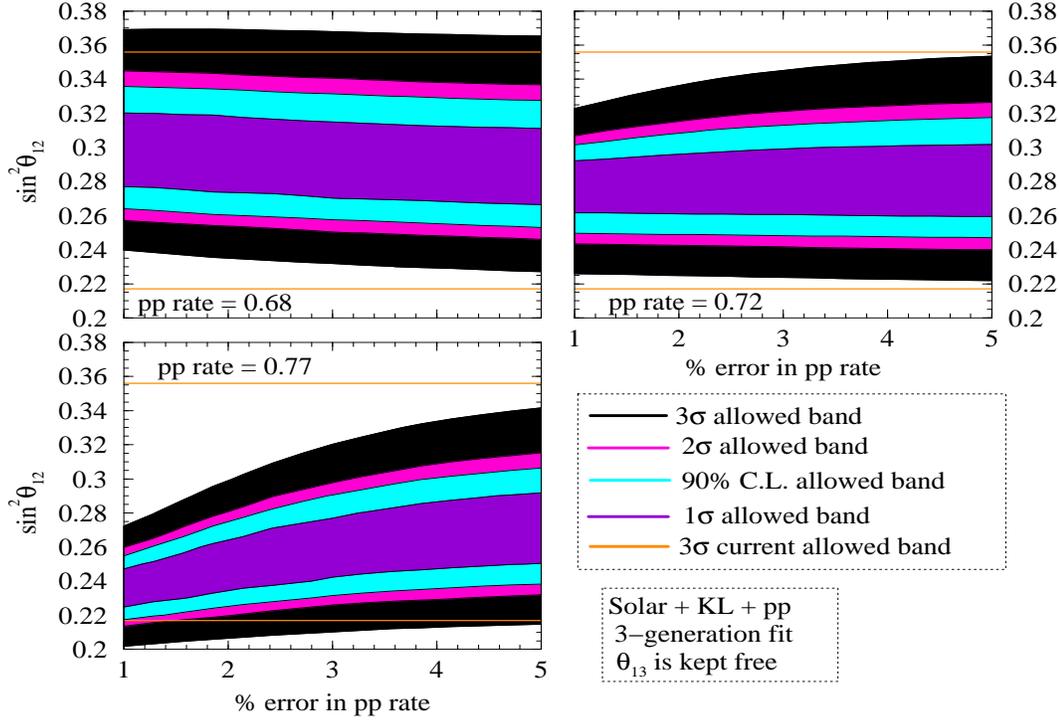}
\caption{\label{fig7}
Same as Figure \ref{fig3} but for a three generation analysis 
including all neutrino data and keeping $\sch$ free.
We also show the 3$\sigma$ allowed band 
of $\sss$ from the current data.
}
\end{figure}
\begin{figure}[t]
\includegraphics[width=14.0cm, height=10cm]
{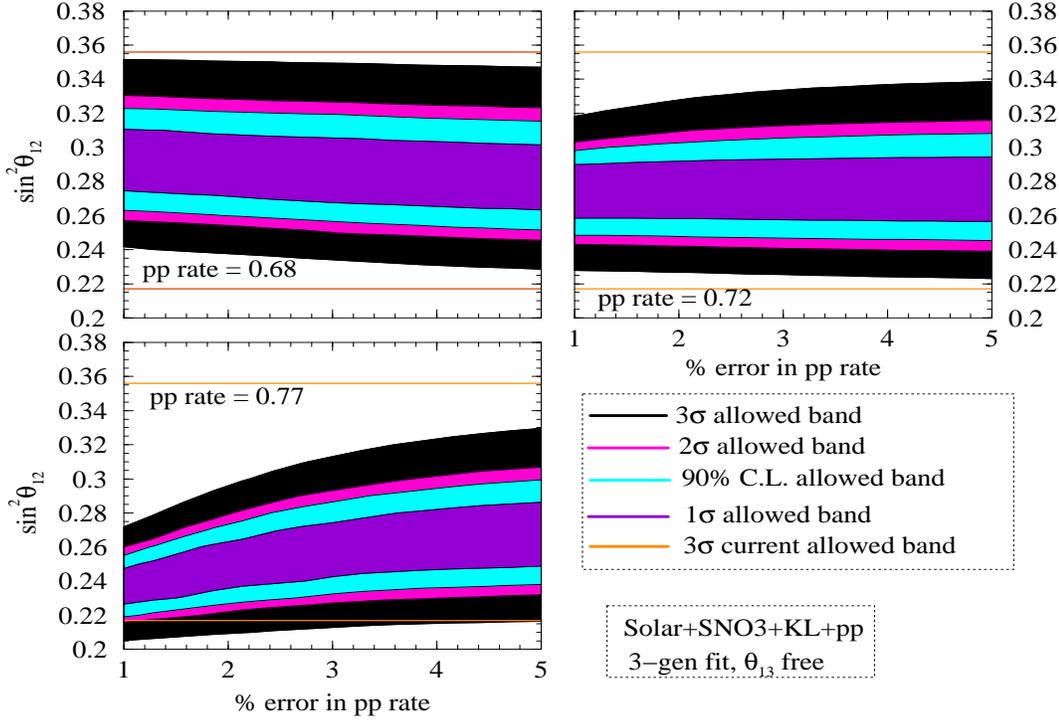}
\caption{\label{fig8}
Same as Figure \ref{fig7} but with 
the projected error from the
third phase of the SNO experiment included.
We show the 3$\sigma$ allowed band from current data. 
}
\end{figure}

   The solar and atmospheric neutrino, K2K and \kl
data suggest the existence of 3-neutrino mixing 
and oscillations (see, e.g.,  \cite{3numix}). 
Actually, all existing neutrino 
oscillation data, except the data of LSND experiment  
\footnote{In the
LSND experiment indications 
for oscillations
$\bar \nu_{\mu}\to\bar \nu_{e}$  
with $(\Delta m^{2})_{\rm{LSND}}\simeq 
1~\rm{eV}^{2}$ were obtained. 
The  LSND results are being tested 
in the MiniBooNE experiment
\cite{MiniB}.}
\cite{LSND}, can be described assuming 
3-neutrino mixing. 
This warrants a 3-neutrino oscillation
analysis of the potential 
sensitivity of a LowNu $pp$
neutrino experiment to $\sss$.

   The 3-neutrino oscillations of interest 
are characterized by the neutrino mass-squared
differences which drive the solar and 
atmospheric neutrino oscillations,
$\Delta m^2_{21} = \Delta m^2_\odot > 0$ and
$\Delta m^2_{31} = \Delta m^2_{atm}$ respectively,
and by the 3 mixing angles
in the Pontecorvo-Maki-Nakagawa-Sakata
(PMNS) neutrino mixing matrix,
$\theta_{12}$, $\theta_{23}$ and $\theta_{13}$.
In the standard parametrization of the
PMNS matrix (see, e.g., \cite{3numix}), 
the angles $\theta_{12}$ and $\theta_{23}$ 
coincide with the angles which 
control the oscillations of
solar and atmospheric neutrinos,
while $\theta_{13}$ is the angle 
limited by the data from 
the CHOOZ and Palo Verde experiments
\cite{chooz}.
The precise limit on $\theta_{13}$
depends strongly on $\Delta m^2_{atm}$ (see e.g. \cite{sgnu04})
The existing atmospheric and reactor 
neutrino data imply \cite{kl766us,kl766others}
\be
\sin^2\theta_{13} < 0.05,~~~~~99.73\%~{\rm C.L.} 
\ee
The aim of the analysis which follows
is to quantify the uncertainty
which the absence of precise knowledge
of the value of the CHOOZ angle $\theta_{13}$
introduces in the precision of 
$\sss$ determination.
%

  The analyses of the latest SK atmospheric 
neutrino and of the global solar neutrino 
data show also that 
\cite{kearns04,kl766us,kl766others}
$|\Delta m^2_{31}| = |\Delta m^2_{{\rm atm}}|
\sim  2.1\times 10^{-3}~{\rm eV^2}$ and
$\Delta m^2_{21} = \Delta m^2_{\odot} 
\sim 8\times 10^{-5}~{\rm eV^2}$. Thus,
we have $\Delta m^2_{21} << |\Delta m^2_{31}|$.
Under this condition the 
probabilities of survival of the 
solar $\nu_e$ and of the reactor 
$\bar{\nu}_e$, relevant for the
3-neutrino oscillation interpretation 
of the solar neutrino and \kl data, 
have the form:
\be
P_{ee}^{3\nu} \cong \cos^4 \theta_{13} P_{ee}^{2\nu} + \sin^4\theta_{13}
\label{3genpee}
\ee
where $ P_{ee}^{2\nu}$ is the
corresponding probability
of survival of $\nu_e$
or $\bar{\nu}_e$ in the case of
2-neutrino mixing (see, e.g., \cite{SP3nuosc88}).
For the reactor $\bar{\nu}_e$ detected at \kl, 
$ P_{ee}^{2\nu}$ is given by eq. (\ref{eq:klpee}). 
For solar neutrinos,
$ P_{ee}^{2\nu} \equiv P_{ee\odot}^{2\nu}$
is the $\nu_e$ survival probability
in the case of 2-neutrino oscillation
\cite{SP88,SPJRich89} in which
the solar electron number density $N_e$
is replaced by $N_e \cos^2\theta_{13}$. From 
eqs. (2), (5), (12) and (13) we get:
\be
P^{3\nu}_{ee}({\br}) &\cong& \cos^4 \theta_{13} \sin^2\thsol, 
\label{pee8b3g} 
\\ 
P^{3\nu}_{ee}({pp}) &\cong& \cos^4 \theta_{13}
( 1 - \frac{1}{2}\sin^22\theta_{12}).
\label{peepp3g}
\ee
%
Note that for a given value of $P^{3\nu}_{ee}({pp})$,
a non-zero value of $\theta_{13}$
{\it decreases} the measured value of $\theta_{12}$, for the 
low energy $pp$ flux. 
On the other hand, for a 
given value of $P^{3\nu}_{ee}({\br})$,
a non-zero $\theta_{13}$
{\it increases} the measured value of $\theta_{12}$, for the 
higher energy $\br$ flux. 

 For $\Delta m^2_{21} << |\Delta m^2_{31}|$,
the probability relevant for the interpretation of 
data from the CHOOZ experiment is given by
\be
P_{eeCHOOZ}^{3\nu} \approx 1 - \sin^22\theta_{13}
\sin^2(\Delta m_{31}^2 L/4E).
\ee
Note that the probability $P_{eeCHOOZ}^{3\nu}$
depends on $\Delta m^2_{31}$,
unlike the probabilities
relevant for the interpretation 
of the solar and \kl data.
In the analysis which we have performed
$|\Delta m^2_{31}|$ was allowed to vary freely 
within  the $3\sigma$ allowed range given in \cite{kearns04}
~\footnote{Details of our 3-neutrino oscillation 
analysis can be found in \cite{kkthree}.}. 

    In Fig. \ref{delchi2g+3g} we show 
the dependence of $\Delta \chi^2$ on $\sin^2\theta_{12}$, 
obtained in a 3-neutrino oscillation analysis 
of the combined KamLAND, CHOOZ and solar neutrino 
data, including the simulated data on the $pp$ neutrino flux.
The results shown are for the 
three illustrative central values of 
$R_{pp}$ considered earlier,
and for the case of 1\% error in $R_{pp}$.
Except for $\sss$, all the other parameters, 
including $\sin^2\theta_{13}$, were 
allowed to vary freely in the analysis. 
For comparison, results 
for $\sin^2\theta_{13}=0$ 
are shown in the same figure.   

We find that 
\begin{itemize}
\item
For a relatively low value 
of the measured $pp$ rate, $R_{pp}= 0.68$,  
a non-zero $\sin^2\theta_{13}$  
leads to smaller minimal and  maximal
allowed values of $\sss$.  
\item
For $R_{pp} =0.72$, the minimal allowed value of $\sss$ 
diminishes, while the maximal allowed value remains unaffected. 
\item
If $R_{pp}=0.77$, both values are practically unaffected. 
\end{itemize}

   To help explain these features, we plot
in Fig.  \ref{probpp} the
$pp$ rate (cf. Eqs. (\ref{rppes}) and (\ref{peepp3g}))
as a function of $\sin^2\theta_{12}$ for 
$\sin^2\theta_{13}=0$ and $\sch=0.05$. 
The three horizontal lines in the figure correspond 
to the three central values of $R_{pp}$
considered in our analysis. 
We see from the figure that,
\begin{enumerate}
\item
For a given value of $\sss$,
a non-zero $\sch$ always reduces the predicted $pp$ rate,
\item
For a given measured $pp$ rate, a non-zero $\sch$ would 
reduce the measured value of $\sss$,
\item
For a given measured $pp$ rate
there is always a limiting value of $\sss$, 
such that
for $\sss$ exceeding this value a non-zero 
$\sin^2\theta_{13}$
does not affect
the
allowed values of $\sss$. 
\end{enumerate}
Points (1) and (2) imply, in particular, that
if for a given value of $\sss$ 
the $pp$ rate predicted assuming 2-neutrino 
oscillations is larger than 
the measured $pp$ rate, a non-zero $\theta_{13}$ 
can improve the quality of the fit. 
Thus, values of $\sss$ smaller 
than the minimal allowed one
determined in a 2-neutrino 
oscillation analysis,
could become allowed in the case 
of 3-neutrino oscillations 
due to non-zero $\sch$.
As a consequence of point 3 we can conclude that
if for a particular value of $\sss$, 
the $pp$ rate predicted in the case of 
2-neutrino oscillations
is larger than the measured $pp$ rate, 
a $\sch$ which differs from 0 substantially
and further lowers the $pp$ rate, 
would not be favored by the data. 
{{We use these features  to explain Fig. \ref{delchi2g+3g}.
However it is to be borne in mind that Fig. \ref{probpp} 
contains only the pp rates while in Figure \ref{delchi2g+3g}
we show the  
results from a combined analysis of global solar and \kl data. 

\begin{itemize} 
\item
We see from Fig. \ref{probpp} that 
the $pp$ rate, predicted in the case 
of 2-neutrino oscillations, 
exceeds 0.72  for $\sss \ltap 0.27$. 
Therefore in this case 
a non-zero $\sch$ improves the 
quality of the fit 
for all values of $\sss \ltap 0.27$.
Thus, relatively small values of $\sss$ 
which are disfavored by the 2-neutrino 
oscillation analysis, could become allowed 
if $\sch\neq 0$. This would lead to a smaller 
minimal allowed value of $\sin^2\theta_{12}$. 
For $\sss > 0.27$, the 
predicted 2-neutrino oscillation 
rate is already lower than 0.72, 
and therefore a non-zero $\theta_{13}$ 
would not change the 
maximal allowed value of $\sss$.

\item
The predicted $R_{pp}$ 
for 2-neutrino oscillations
can be greater than $R_{pp}= 0.77$
only for relatively small values of 
$\sss$, which are strongly disfavored 
(if not ruled out) by the current data.
Consequently, a non-zero $\sch$ 
is not expected to make any impact 
on the $\sss$ determination if 
the measured $R_{pp}= 0.77$.

\item 
In the case of a measured $R_{pp}=0.68$,
the predicted 2-neutrino oscillation 
$pp$ rate is larger than 0.68
for $\sss \ltap 0.34$
and a non-zero value of $\sch$ can 
improve the quality of the fit for a  
large range of values of $\sss$.
This explains why the minimal allowed value of $\sss$ 
diminishes considerably for $\sch \neq 0$
in Fig. \ref{delchi2g+3g}. 
The maximal allowed value of $\sss$ is also seen 
to reduce.  
The reason for this can be understood 
if one notes that 
the values of $\sss$
favoured by the $pp$ experiment,
for $\sin^2\theta_{13} =0$,
would
correspond to a relatively higher
value of $R_{CC}/R_{NC}$ compared to that  measured at SNO. 
If $\theta_{13}$ is non-zero, then the same $R_{pp}$ could be
produced at a lower value of $\theta_{12}$.
Since the predicted
$R_{CC}/R_{NC}$ in SNO is given by Eq. (\ref{pee8b3g}),
this lower $\sss$ coupled with non-zero $\theta_{13}$
would help
reduce the $^8B$ probability
and the
$pp$ ``data'' could be
``reconciled'' with the SNO CC/NC data.
Therefore for
$R_{pp}= 0.68$,
the best-fit from a three generation analysis comes at a non-zero value of
$\sin^2\theta_{13}$ and a lower
value of $\sin^2\theta_{12}$.
Note that the value of the global
$\chi^2_{min}$ for the three generation case ($\theta_{13}\neq 0$)
is lower
than the two-generation case ($\theta_{13}=0$).
However as noted before, since the predicted $R_{pp}$ is already
lower than 0.68 for $\sss \gtap 0.34$, the $pp$ experiment would
force $\theta_{13}=0$ for these high values of $\sss$.
Thus above $\sss \gtap 0.34$
the $\chi^2$ remains the same irrespective of whether $\theta_{13}$
was kept free (three-generation) or fixed at zero (two-generation).
However, since the global $\chi^2_{min}$ was lower for the three-generation
fit, the very high values of $\sss$, which were allowed in the
two-generation analysis get disfavored by the three-generation fit.

\end{itemize} 
}}
 
  
  In Fig. \ref{fig7} we depict 
the $\sin^2\theta_{12}$ sensitivity  
of the world neutrino oscillation 
data, including the sample data on the $pp$ rate,
as a function of \% error in the measured $pp$ rate   
in the case when $\theta_{13}$ is kept free. 
In Table \ref{pp3gspread} we give the corresponding  
3$\sigma$ ranges of allowed values and spread of $\sss$
(columns 3 and 4). 
A comparison of this figure with Fig. \ref{fig3} 
and the Table \ref{pp3gspread} with Table \ref{pp2gspread}
demonstrates clearly all the specific features
associated with the three values of the $pp$ rate
discussed above: 
\\
For  $R_{pp}=0.77$, $\theta_{13} \neq 0$ 
hardly makes any difference. 
\\
For $R_{pp}=0.72$, 
the minimal allowed value of $\sss$ diminishes
as a consequence of $\theta_{13} \neq 0$,
while the maximal allowed value is unaltered. 
\\
For $R_{pp}=0.68$, the inclusion of $\theta_{13} \neq 0$
in the analysis leads to 
a reduction of both the minimal 
and maximal allowed values of
$\sss$. 
\\
For  $R_{pp}=0.72$ and 0.68  
cases we also note that the minimal
value of $\sin^2\theta_{12}$ increases as the error in $R_{pp}$ increases
and the SNO data begins to have a greater influence on the fit.''
Thus,   
the effect of the uncertainty due to $\sch$ on 
the precision of $\sss$ measurement 
decreases as the error in the $pp$ rate increases.

Figure \ref{fig8} shows 
the corresponding sensitivity 
plot of $\sss$ from  
a 3-neutrino oscillation analysis of 
the global neutrino oscillation data 
including both the (hypothetical) $pp$ 
rate data and the prospective data (and errors)
from phase-III of the SNO experiment. 
The corresponding allowed ranges and spread of $\sss$ is given in 
columns 5 and 6 of Table \ref{pp3gspread}. 


From the expression of the
3-neutrino oscillation probability,
eq. (\ref{3genpee}), we see that
the factor $\cos^4\theta_{13}$ acts
like a ``normalization constant''.
Since the current $3\sigma$ limit
on this parameter is $\sch < 0.05$,
one would get a $\sim 10\%$
uncertainty in $P_{ee}^{3\nu}$
and would expect similar uncertainty
to appear in the value of $\sss$
determined using the $pp$ rate.
The actual increase
in the $\sss$ uncertainty due to $\sch$
is smaller than $\sim 10\%$,
typically being $\sim 3\%$ for the
plausible values of the $pp$ rate ($R_{pp} = 0.72$)
we have considered.
Even for the limiting case of
$R_{pp}=0.68$, the
maximal increase is by
$\sim 6\%$


The main reason for this is that
when we include both "low" and "high" energy experiments
in the global analysis
there are two conflicting trends.
While
a non-zero value of $\sch$
would have a tendency to
lower the value of $\sss$ determined
from a 2-neutrino oscillation
analysis of the data from the $pp$ experiment,
but it would also have a tendency to increase the
value of $\sss$ determined from the data of SNO and SK
(i.e., \br neutrino) experiments. 
At the lower bound, in general, 
$pp$ tries to shift the fit to non-zero $\sch$ and 
hence lower values of $\sss$ but the global data including SNO 
prevents that.  
On the other hand, at the upper bound, in general,  
SNO can push the fit to $\sch \neq 0$ and higher $\sss$ but 
$pp$ prefers to keep it at 
$\sch=0.0$ which corresponds to a lower maximal allowed value 
of $\sss$.

{\it The uncertainty in the value of $\sch$ leads to 
an error in $\sss$ at the  
few percent level only when $\sss$
is determined using data on the $pp$ rate 
together with the global solar 
and reactor neutrino data.}
Nevertheless,
the 
spread in the value of $\sss$
remains well above 12.5\% and 
typically exceeds 16\% at $3\sigma$.
As we will show in the next section,
$\sss$ could be measured
with a considerably higher precision 
in a reactor neutrino experiment with a baseline 
tuned to SPMIN. 

\begin{table}[htb]
\begin{center}
\begin{tabular}{||c||c||c|c||c|c||} 
\hline \hline
& & \multicolumn{2}{c||}{\rm solar+reactor +pp} & \multicolumn{2}{c||} 
{\rm solar(SNO3)+reactor+pp} \\ \cline{3-6} 
pp rate &\% error &
3$\sigma$ range & spread & 3$\sigma$ range & spread \\ \hline
0.68 & 1 & 0.24 - 0.37 & 21.3\% & 0.24 - 0.35 & 18.6\%\\
& 2 & 0.23 - 0.37 & 23.3\% & 0.24 - 0.35 & 18.6\% \\
& 3 & 0.23 -0.37 & 23.3\% & 0.23 - 0.35 & 20.7\%  \\
& 4 & 0.23 - 0.37 & 23.3\% & 0.23 - 0.35 & 20.7\%  
\\ \hline
0.72 & 1 & 0.225 - 0.32 & 17.4\% & 0.23 - 0.32 & 16.4\% \\
& 2 & 0.22 - 0.34 & 21.4\% & 0.23 - 0.33 & 17.9\% \\
& 3 & 0.22 - 0.34 & 21.4\% & 0.22 - 0.33 & 20.0\% \\
& 4 & 0.22 - 0.35 & 22.8\%  & 0.22 - 0.34 & 21.4\% 
\\ \hline
0.77 & 1 & 0.2 - 0.27 & 14.9\% & 0.20 - 0.27 & 14.9\% \\
{}   & 2 & 0.21 - 0.30 & 17.6\% & 0.21 - 0.29 & 16.0\% \\
& 3 & 0.21 - 0.32 & 20.8\% & 0.21 - 0.31 & 19.2\% \\
& 4 & 0.21 - 0.33 & 22.2\% & 0.21 - 0.32 & 20.8\% \\
\hline 
\end{tabular}
\end{center}
\caption{
The 3$\sigma$ allowed ranges and \% spread  of 
$\sin^2\theta_{12}$, 
obtained from a 3-neutrino oscillation 
analysis of the global solar and reactor neutrino data, 
including the hypothetical data on the $pp$ rate.
}
\label{pp3gspread}
\end{table}
\section{Measuring $\theta_{12}$ in a Reactor Experiment at SPMIN}

   In this section we investigate 
the possibility of measuring the 
solar neutrino mixing parameter $\sss$ 
in a reactor $\bar{\nu}_e$
oscillation experiment,
in which the baseline is chosen
to correspond to a
minimum of the $\bar{\nu}_e$
survival probability
(SPMIN) \cite{th12}. 
We will consider in what follows 
an experiment similar to \kl, 
but with a baseline tuned to the SPMIN. 
The condition of SPMIN reads
\be
{L_{min} \approx 1.24 \frac{E/MeV}{\dm/eV^2}}~m
\label{Lspmin}
\ee
For the ``old'' low-LMA best-fit value 
of $\ms=7.2\times 10^{-5}$ eV$^2$, 
the baseline which allows the most 
accurate measurement of 
$\sss$ was found to be 70 km \cite{th12}. 
For these $\ms$ and baseline
the SPMIN appears at the prompt $e^+$ 
energy of $E_{vis}=3.2$ MeV
\footnote{For the details of our reactor code and
statistical analysis see 
refs. \cite{th12,th12hlma,skgd}.},
$E_{vis} \cong E - 0.8$ MeV.
The latter corresponds to the
maximum of the $e^+$ (event) spectrum
in the absence of oscillations.
Obviously, the energy
 $E_{vis} \sim 3.2$ MeV,
is the most relevant for
the statistics of the experiment.
We will show in this section that for 
the current global best-fit value of 
$\ms=8.3\times 10^{-5}$ eV$^2$, 
$\sss$ could be measured with an accuracy of 
$\sim 2\%~(6\%)$ at $1\sigma~(3\sigma)$
if the baseline chosen is $L\sim 60$ km.
For $\ms=8.3\times 10^{-5}$ eV$^2$ and
$L\sim 60$ km, the SPMIN 
appears at $E_{vis}=3.2$ MeV in the $e^+$ spectrum.  
We extend our earlier work 
\cite{th12,th12hlma,skgd}
by investigating in detail the dependence 
of the precision of $\sss$ measurement on  
the true value of $\ms$,
the baseline, the statistics 
and on the systematic error 
of the experiment. 
We obtain results assuming 
2-neutrino oscillations
and compare them with
the results of a 3-neutrino 
oscillation analysis.
In the latter $\sch$ is allowed to vary 
freely within its currently allowed range. 
We discuss also the relevance of the geo-neutrino 
flux for the precision of $\sss$
measurement.
\begin{figure}[t]
\includegraphics[width=14.0cm, height=10cm]{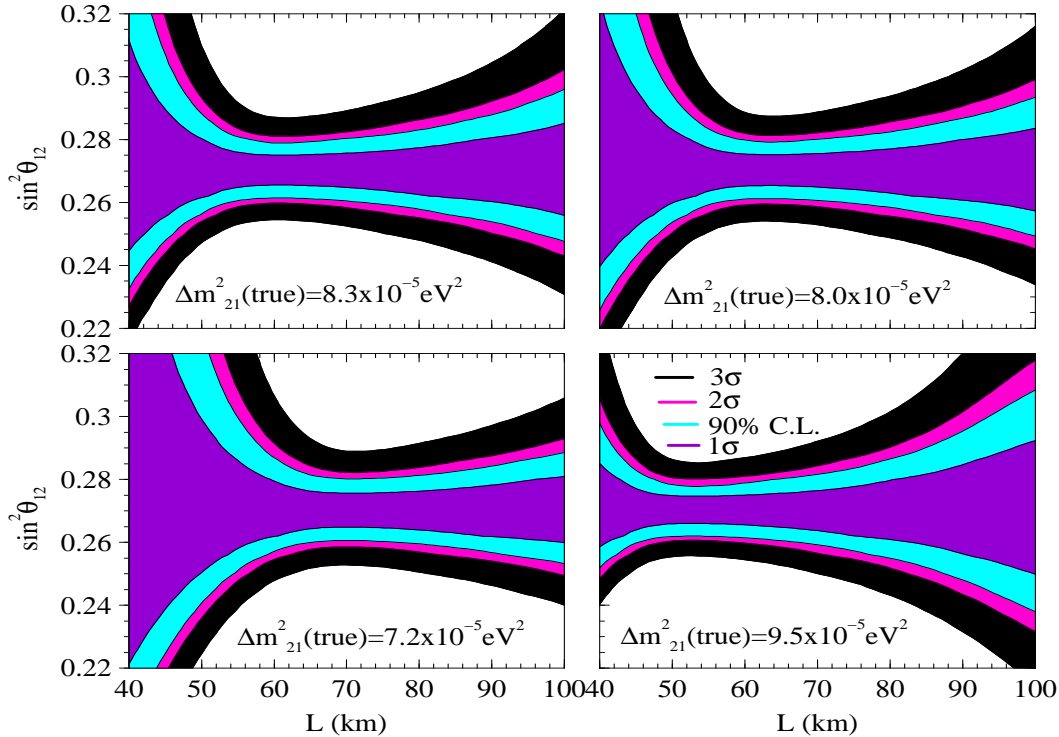}
\caption{\label{reactL}
Sensitivity plots showing the $1\sigma$,  $1.64\sigma$, $2\sigma$, 
and $3\sigma$ range of allowed values for $\sss$ 
as a function of the baseline $L$. The 4 panels are for 4 different 
true value of $\ms$. The true value of $\sss$ is assumed to be
0.27 in all the cases. The $\ms$ is allowed to vary freely in the fit.
}
\end{figure}
\begin{figure}[t]
\includegraphics[width=14.0cm, height=10.0cm]{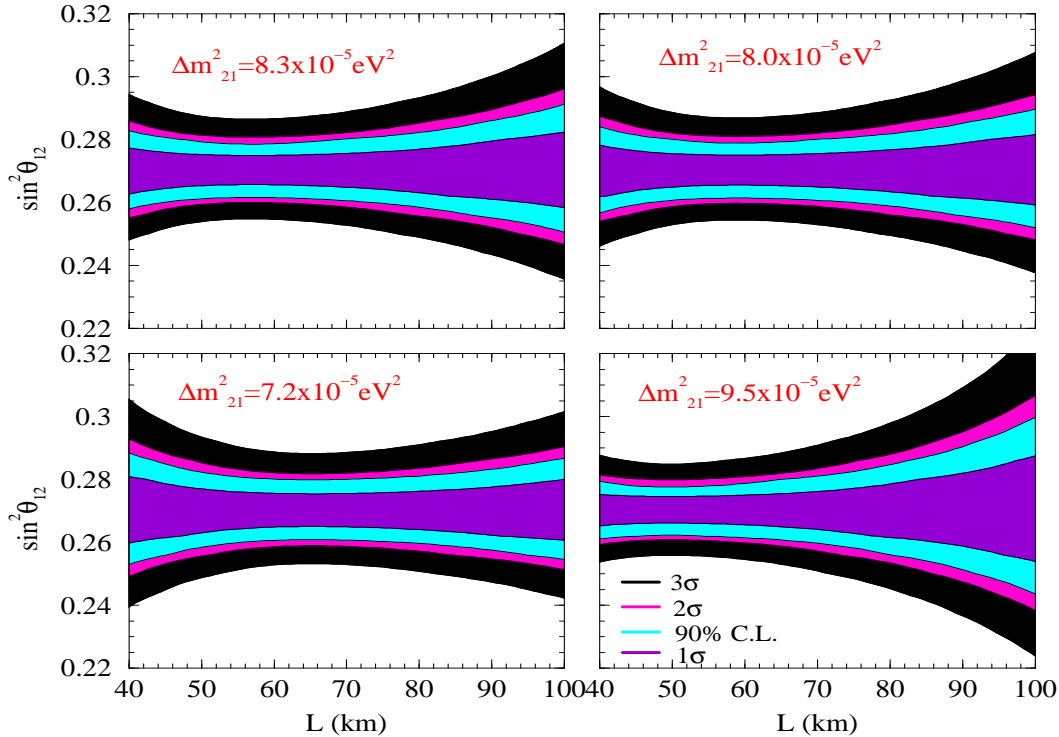}
\caption{\label{reactLfixedd21}
The same as in Fig. \ref{reactL}, but 
for fixed $\ms$ 
having an assumed true value indicated
in each of the panels of the figure.
}
\end{figure}
\begin{figure}[t]
\includegraphics[width=14.0cm, height=10.0cm]{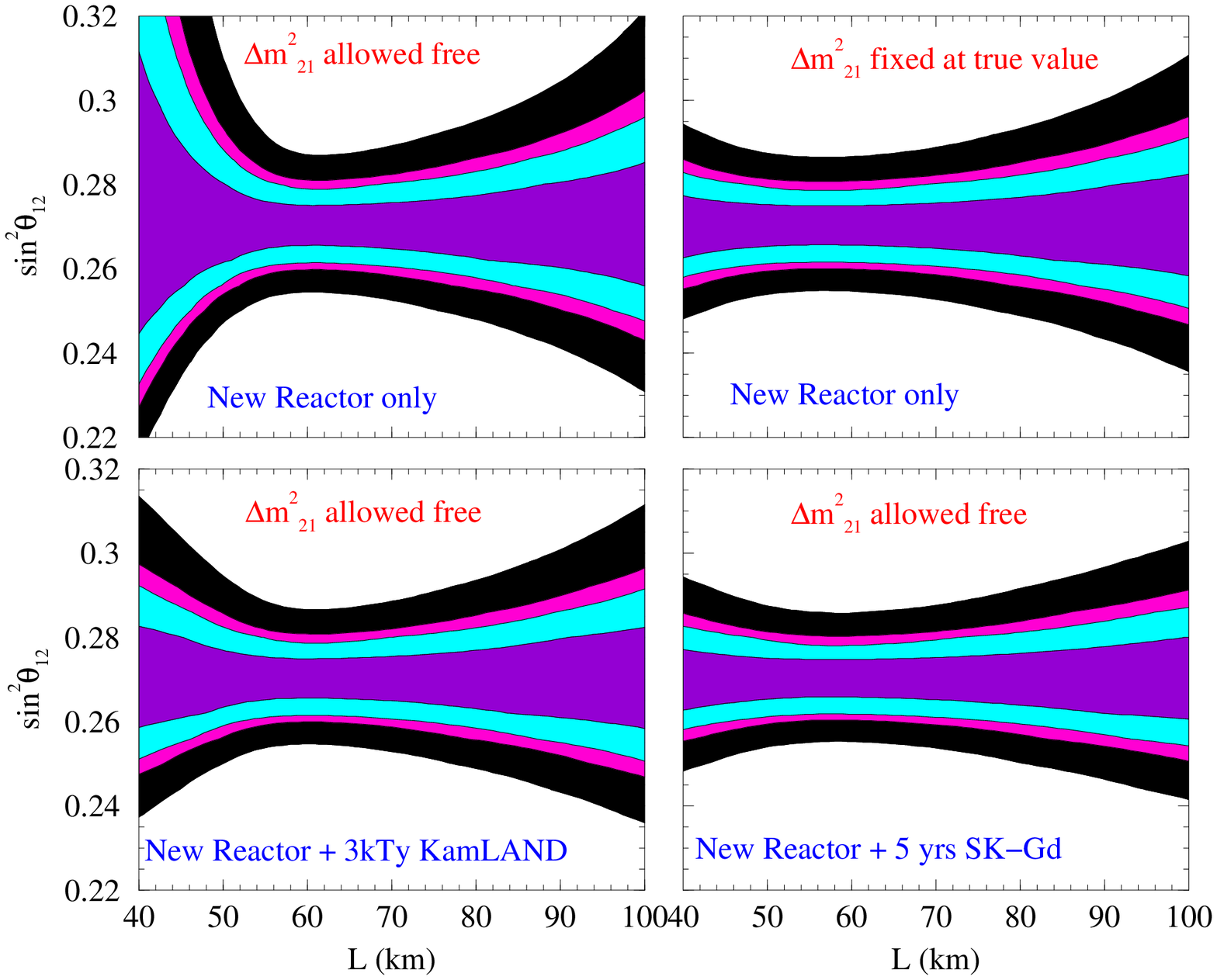}
\caption{\label{reactLdelkl}
Sensitivity plots showing the $1\sigma$,  $1.64\sigma$, $2\sigma$, 
and $3\sigma$ range of allowed values for $\sss$ 
as a function of the baseline $L$. The true values of $\ms(true)=
8.3\times 10^{-5}$ eV$^2$ and of $\sss(true)=0.27$ for all panels.
The upper left hand (right hand) panel is 
obtained allowing $\ms$ to vary freely
(for fixed $\ms(true)= 8.3\times 10^{-5}$ eV$^2$).
The results shown in the 
lower left hand (lower right hand) panel are from
combined analysis of 3 kTy data from KamLAND
(5 year data from the suggested SK-Gd experiment)
and 73 GWkTy data from a SPMIN reactor 
experiment.
}
\end{figure}
\subsection{Sensitivity to $\sss$ and the Baseline of the Experiment}

  In Fig. \ref{reactL} we show 
the sensitivity to  $\sss$, expected 
in a reactor experiment, as a function 
of the baseline $L$.
We assume a total systematic uncertainty of
2\% and consider statistics of 73 GWkTy 
(given as a product of reactor power in GW and 
the exposure of the detector in kTy).
The detector material composition 
is assumed to be the same  as that 
of KamLAND and so the detector 
considered has the 
same number of target protons 
per kton as KamLAND. 
The total reactor power, the 
detector size and the exposure time 
are kept the same for all baselines.
Thus, for longer baseline 
the number of events would 
decrease as $\sim L^{-2}$.  
We assume that the true value of 
$\sss=0.27$ and simulate the 
prospective observed positron spectrum 
in the detector for four different assumed 
true values of $\ms$.  This figure
is obtained for $\sch=0$. 

  We define a $\chi^2$ function given by
\be
\chi^2 =  \sum_{i,j}(N_{i}^{data} - N_{i}^{theory})
(\sigma_{ij}^2)^{-1}(N_{j}^{data} - N_{j}^{theory})~,
\label{chig}
\ee
where $N_i^\alpha$ $(\alpha=data,theory)$ 
is the number of events in 
the $i^{\rm th}$ bin, $ \sigma_{ij}^2$ 
is the covariant error matrix 
containing the statistical and systematic 
errors and the sum is over all bins. 
We use this $\chi^2$ to fit the 
simulated spectrum data and get the 
``measured'' value of $\sss$, keeping $\ms$ free.
We simulate the spectrum at each 
baseline and plot the range of values 
of $\sss$ allowed by the 
simulated data
as a function of the baseline. 
The baseline at which the band 
of allowed values of $\sss$ is 
most narrow is the ``ideal'' baseline 
for the SPMIN reactor experiment.
The figure confirms that 
this ``ideal'' baseline depends critically
on the true value of $\ms$
(cf. eq. (\ref{Lspmin})). 
The optimal baseline for 
the true value of 
$\ms=8.3~(8.0)\times 10^{-5}$ eV$^2$ is 
seen from Fig. \ref{reactL} to be $60~(63)$ km, 
while for the ``old'' low-LMA best-fit value 
of $\ms=7.2\times 10^{-5}$ eV$^2$ the best baseline 
would be 70 km. At the optimal baseline the SPMIN 
reactor experiment can achieve an unprecedented
accuracy of $\sim 2\%~(6\%)$ at $1\sigma~(3\sigma)$ in 
the measurement of $\sss$.

   Figure \ref{reactL} suggests
that the optimal baseline 
for a given true value of $\ms$ is very 
finely tuned. For instance, 
if for $\ms(true)=8.3 \times 10^{-5}$ eV$^2$
we change the baseline from $L=60$ to $L=50$ km, the 
sensitivity in $\sss$ decreases from $\sim 2\%~(6\%)$ 
to $\sim 3\%~(11\%)$ at $1\sigma~(3\sigma)$.
However, note that Fig. \ref{reactL} 
was obtained by allowing $\ms$ to vary freely. 
This is equivalent to assuming that
both $\sss$ and $\ms$ are
determined in the reactor SPMIN experiment.
For some baselines, especially at smaller $L$, 
the oscillation induced spectral 
distortion is not large enough to measure 
$\ms$ sufficiently accurately, 
while for the longer baselines 
the statistics is lower. These factors lead 
to a certain uncertainty 
in the determination of $\ms$ with
the experimental set-up under discussion. 
The uncertainty in the 
$\ms$ determination translates into additional
uncertainty in the measured $\sss$. 
If $\ms$ could be 
measured with a sufficiently high precision in
an independent experiment, the uncertainty 
in $\sss$ due to $\ms$ would be reduced.

    Figure \ref{reactLfixedd21} represents 
a sensitivity plot similar to that shown
in Fig. \ref{reactL}, but obtained for  $\ms$ 
fixed at its assumed true value 
(indicated on each of the panels). 
As Fig. \ref{reactLfixedd21} shows,
for fixed $\ms$ 
assumed to have been determined
with a sufficiently high precision
in an independent experiment, 
the choice of the baseline for 
setting up the SPMIN experiment becomes broader.  
It follows from Fig. \ref{reactLfixedd21} that 
for $\ms(true)=8.3 \times 10^{-5}$ eV$^2$,
for instance, the change of the baseline 
from $L=60$ to $L=70$ km, 
leads to a minor increase of 
the uncertainty in the value of $\sss$ 
from $6.1\%$  to $6.3\%$ at $3\sigma$. 

   It is actually quite possible
that $\ms$ will be measured with 
a rather high accuracy in the future. 
The \kl experiment could 
determine $\ms$ with an error of about 7\% 
(at 3$\sigma$) using data of 3 kTy  
\cite{prekl,th12,kl766us}. 
The suggested SK-Gd experiment \cite{gadzooks}
has the potential of measuring the 
value of $\ms$ with an error of  $\sim 2-3$\% (at 3$\sigma$) 
\cite{skgd}.  In Fig. \ref{reactLdelkl} we show the 
sensitivity to $\sss$ expected if we combine the SPMIN reactor 
data with 3 kTy prospective data from \kl (lower left panel) and 
simulated 5 year data from the suggested 
SK-Gd experiment (lower right panel). The upper panels 
were obtained using data from the SPMIN reactor experiment alone. 
For all the panels we have assumed 
$\ms(true)=8.3 \times 10^{-5}$ eV$^2$.
In the  upper left panel we allow 
$\ms$ to vary freely, while in the 
upper right panel $\ms$ is fixed 
at the assumed true value. 
We note that if the SPMIN reactor 
data is combined with 5 year data from the SK-Gd experiment, 
the choice of optimal baseline is much wider
since $\ms$ would be determined with a relatively high 
precision by the SK-Gd experiment. 
With the addition of the SK-Gd results to the 
total data set, the spread in $\sss$ 
is $\sim 5.7\%$ at $3\sigma$. 
The combined SPMIN reactor and \kl 3 kTy data would 
yield an uncertainty in the value of $\sss$ of 
$\sim 5.9\%$  at $3\sigma$. Since the analysis of the 
combined \kl (or SK-Gd) and SPMIN reactor data  
confirms that the effect of $\ms$ on 
the $\sss$ sensitivity can be negligible, 
we will take $\ms$ to be fixed 
for the remainder of this section.
\subsection{Impact of Statistical and Systematic Errors
on $\sss$ Sensitivity}

\begin{figure}[t]
\includegraphics[width=14.0cm, height=10.0cm]{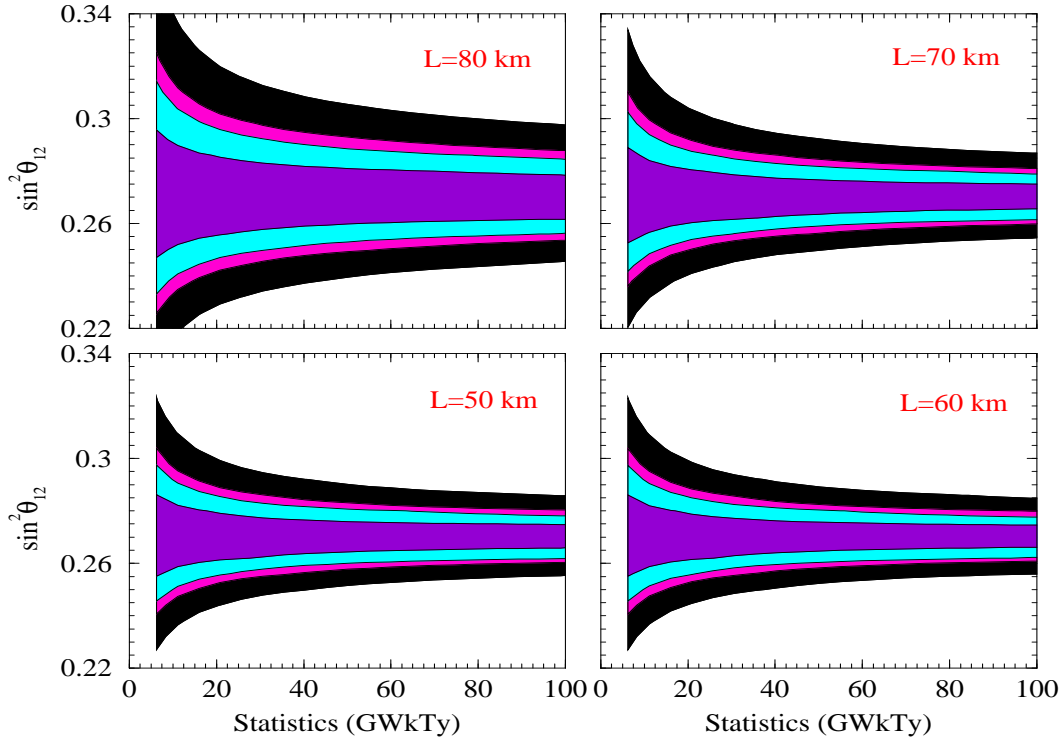}
\caption{\label{reactstat}
Sensitivity plots showing the $1\sigma$,  $1.64\sigma$, $2\sigma$, 
and $3\sigma$ ranges of allowed values for $\sss$ 
as a function of the statistics in units of GWkTy. 
All the four panels correspond to a fixed 
value of $\ms=8.3\times 10^{-5}$ eV$^2$. 
}
\end{figure}
\begin{figure}[t]
\includegraphics[width=14.0cm, height=10.0cm]{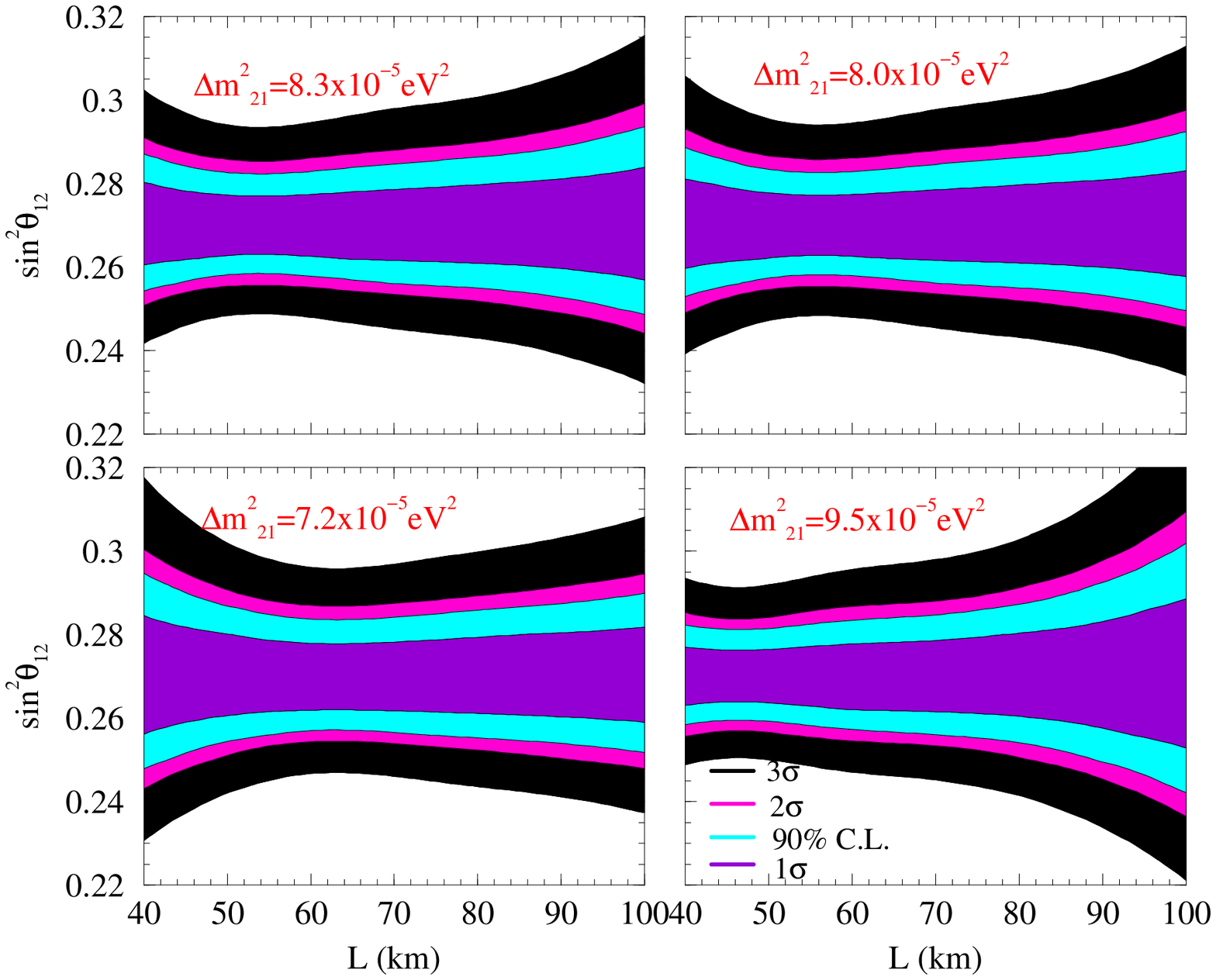}
\caption{\label{reactsyst}
The Same as in Fig. \ref{reactLfixedd21} but for a systematic uncertainty
of 5\%.
}
\end{figure}

  One of the important requirements for the 
type of high precision experiment we are discussing
is the accumulation of relatively high statistics
in a reasonable period of time. 
Since the statistics falls 
as $\sim L^{-2}$
and since rather long baselines are required
for a precision measurement of the solar 
neutrino oscillation parameters, 
for a given reactor power 
longer baselines would imply
bigger detectors and larger exposure times. 
Thus, the question about the dependence of 
the precision of measurement of $\sss$ in a reactor SPMIN
experiment on the statistics of the experiment 
naturally arises.
In  Fig. \ref{reactstat} we show 
the effect of the statistics 
on the $\sss$ sensitivity in the case 
of $\ms(true)=8.3 \times 10^{-5}$ eV$^2$. 
The four panels are obtained for 
four different sample baselines of 50, 60, 70 and 80 km.
The range of allowed values of $\sss$ 
is shown as a function of the 
product of reactor power and 
the detector mass and exposure time.
For $L=60$ km, for instance, the uncertainty in 
$\sss$ diminishes from 3\% (10\%) to 
2\% (6\%) at $1\sigma(3\sigma)$ as 
the statistics is increased from 20 GWkTy to 60 GWkTy. 
Note that the difference in the 
$\sss$ precision for 60 GWkTy and 73 GWkTy
(used in Figs. \ref{reactL} and \ref{reactLfixedd21}) is 
marginal, and shows up only in the first place in decimal in 
the value of the spread.

  Another important aspect which determines the potential 
of the experiment for precision measurement of $\sss$
is the systematic uncertainty. Obviously, smaller
systematic errors are preferable.
All the plots presented so far in this section 
have been generated with an assumed 2\% systematic error.
The systematic uncertainty in 
the \kl experiment is about 6.5\%. Most of it
comes from the uncertainty in 
the detector fiducial mass and the reactor power.
Our choice of 2\% for the systematic error is 
based on the optimistic assumption
that the  error in the flux 
normalization could be reduced 
sufficiently by using the near-far detector set-up.
One could envisage the $\theta_{12}$ reactor 
SPMIN experiment as a second leg of a 
reactor experiment dedicated to 
measure $\theta_{13}$ (see, e.g., \cite{Whiteth13}). 
The detector for measuring $\theta_{13}$ 
could then effectively be used as near detector 
for the long baseline SPMIN experiment for 
high precision measurement of $\theta_{12}$. 
It should be added that 
the errors due to the
uncertainties in the threshold energy 
and $\bar{\nu}_e$ spectrum
have also to be reduced to achieve the 
systematic error of 2\%.
Experimentally this could be a very challenging task.

   Since systematic uncertainties may be difficult to 
reduce in the experiment under discussion, 
we estimate next how much the precision on 
$\sss$ deteriorates as the systematic error increases.
Figure \ref{reactsyst} shows the effect of increasing the 
systematic error from the 2\% assumed by us 
to the rather conservative value of 5\%. For 
$\ms(true)=8.3 \times 10^{-5}$ eV$^2$,
the spread in $\sss$ at $L=60$ km
increases from 6.1\% to 8.6\% at $3\sigma$, 
as the systematic error is increased from 2\% to 5\%. 
We conclude that the effect of systematic uncertainty on the 
precision of $\sss$ measurement is important, 
but its impact is not dramatic as long as the 
systematic error does not exceed 5\%. Similar conclusion
regarding     the effect of a 4\% systematic error on the 
accuracy of $\sss$ determination in the SADO experiment was
reached in \cite{minat12}.

\subsection{The Uncertainty Due to $\sch$}

\begin{figure}[t]
\includegraphics[width=14.0cm, height=10.0cm]{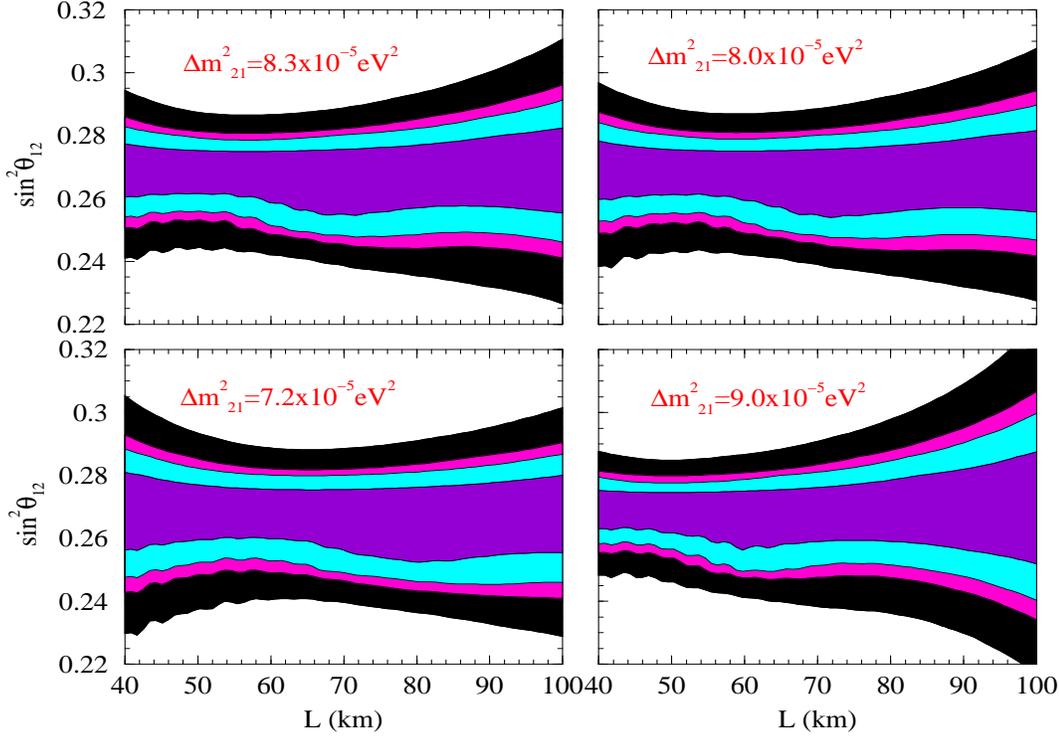}
\caption{\label{reactLth13}
The same as in Fig. \ref{reactLfixedd21} 
but for a 3-neutrino oscillation analysis in which
$\sch$ is allowed to vary freely within its current $3\sigma$
allowed range. 
}
\end{figure}
As we have discussed earlier in connection with the 
KamLAND experiment, the 
3-neutrino oscillation survival probability
for the reactor $\bar{\nu}_e$ of interest is given by
\be
P_{ee}
\approx\cos^4\theta_{13}
\left( 1 - \sin ^{2}2\theta_{12}
\sin^2\frac{ \Delta{m}^2_{21} \, L }{ 4 \, E_{\nu} }\right)~,
\label{pee3g}
\ee 
where the term $\sim \sin^4\theta_{13}$ has been neglected.
Therefore the uncertainty in $\sch$, eq. (12),  
brings up to a $\sim 10\%$ 
uncertainty in the value of the
$\bar{\nu}_e$ survival probability.
Since the factor $\cos^4\theta_{13}$ can only 
reduce the survival probability, 
it does not affect the upper limit 
of the allowed range of $\sss$. However,
it can have an effect on the minimal allowed value 
of $\sss$ reducing it further, and thus 
can worsen, in principle, the precision of the experiment
to $\sss$.

The  {\it additional} 
error on $\sin^22\theta_{12}$ coming from the 
uncertainty in $\sch$ can be roughly estimated 
using eq. (\ref{pee3g}) as \cite{th12hlma},
\be
\delta(\sin^22\theta_{12})\approx 
\frac{2\Delta P_{ee}\sin^2\theta_{13}}
{\sin^2\frac{ \Delta{m}^2_{21} \, L }{ 4 \, E_{\nu} }}
+ 2~\frac{(1 - \sin^2 2\theta_{12}
~\sin^2\frac{ \Delta{m}^2_{21} \, L }{ 4 \, E_{\nu} })
~\Delta (\sin^2\theta_{13})}
{\sin^2\frac{ \Delta{m}^2_{21} \, L }{ 4 \, E_{\nu} }}~,
\label{extra}
\ee
%
\noindent
where $\Delta P_{ee}$ and $\Delta (\sin^2\theta_{13})$ are the 
uncertainties in the determination of the survival probability 
and $\sch$, respectively.
In the SPMIN region we are interested in one has
$\sin^2(\Delta{m}^2_{21}L/ 4E_{\nu})\sim 1$ 
and therefore
\be
\delta(\sin^22\theta_{12})\approx 
2\Delta P_{ee}\sin^2\theta_{13}
+ 
2~\cos^2 2\theta_{12} ~\Delta (\sin^2\theta_{13})~.
\label{extraour}
\ee
%
Thus, for a reactor SPMIN set-up,
the first term gives an extra contribution of 
about $2\Delta P_{ee}\sin^2\theta_{13}$
to the allowed range of $\sin^22\theta_{12}$. 
Even under the most conservative conditions 
one can expect that $\Delta P_{ee} \ltap 0.1$, 
so this term could give an additional contribution 
of $\ltap 0.01$ to the allowed 
range of $\sin^22\theta_{12}$. 
The second term is 
independent of the precision of a given experiment. 
It depends only on the best-fit 
value of $\cos^2 2\theta_{12}$ and on 
the uncertainty in $\sch$.
For the current $3\sigma$ upper limit 
on $\sch$ of 0.05  and 
best-fit value of $\cos^2 2\theta_{12} = 0.19$, 
the second term would lead to an increase 
in the uncertainty in $\sin^22\theta_{12}$ by about 
 0.02 only. The suppression of 
this term is mainly due to the 
presence of the $\cos^2 2\theta_{12}$ factor, 
which is relatively small 
for the current best-fit value. 
Thus, even though the 
current uncertainty in $\sch$ brings a
10\% uncertainty in the 
value of $P_{ee}$, it increases the allowed range 
of $\sin^22\theta_{12}$ only by a few \%,
if one uses a reactor experiment with a 
baseline tuned to the SPMIN for 
a high precision measurement of $\sin^22\theta_{12}$.

  The above conclusions are illustrated in 
Fig. \ref{reactLth13} showing the uncertainty in
$\sss$ expected in the case when $\theta_{13}$ is 
allowed to vary freely in its currently 
allowed range of $\sch < 0.05$.
The figure confirms that the upper 
bound on $\sss$ remains unaffected by the 
$\sch$ uncertainty, while the minimal allowed value
diminishes somewhat, 
increasing the uncertainty in $\sss$. 
However, for the baseline which
corresponds to the SPMIN, 
the sensitivity reduces only by $2-3\%$ 
in spite of the 10\% uncertainty in $\sch$. 
For $\ms(true)=8.3 \times 10^{-5}$ eV$^2$, for instance,
the uncertainty in $\sss$ increases from 6.1\% to 8.7\% at $3\sigma$.
\subsection{On the Impact of Geo-Neutrino Flux}

Our Earth is known to be a huge heat reservoir and is estimated to
radiate about 40 TW of heat. A large fraction ($\sim 16\%$) of 
this is believed to be radiogenic in origin, coming from the 
decay chain of $^{238}U$, $^{232}Th$ and $^{40}K$. 
The radioactive decays of these isotopes produce  antineutrinos  
in the beta decay processes of their decay chains. These 
$\anue$ coming from inside the Earth are usually called  
Geo-neutrinos ($\bar\nu^{geo}_e$) \cite{eder}. 
The maximum energy of
the $\bar\nu^{geo}_e$ produced in the $^{40}K$ decay chain is 
only $E_\anue = 1.31$ MeV, which is below the 
detection energy threshold of 
$\anue$ in scintillation detectors 
of the type of KamLAND we are considering.
However, the $\bar\nu^{geo}_e$ from $^{238}U$ and $^{232}Th$ have 
maximum energy of $E_\anue = 3.26$ MeV and 
$E_\anue = 2.25$ MeV, respectively, and can be 
observed in scintillation detectors. 

 The flux of $\bar\nu^{geo}_e$ is unknown. Even the 
total heat radiated by Earth has a rather large 
uncertainty: it could be $31-40$ TW. 
There is no direct measurement 
of the abundances of the $^{238}U$ and $^{232}Th$ 
inside the Earth. One can estimate their 
abundances using the meteoritic and seismic
data. This results in the $\bar\nu^{geo}_e$ flux 
being largely model dependent and uncertain. 
Most models give 
the bulk $^{232}Th$/$^{238}U$ ratio as 
$^{232}Th$/$^{238}U \sim 3.8$. 
Even the value of this ratio could have a realtively 
large error (e.g., the authors of \cite{flprgeo} 
estimate this error as 14\%).
The measurement of the $\bar\nu^{geo}_e$ 
flux would lead to a better understanding 
of the interior of the Earth, and is therefore 
a very important branch of neutrino 
physics in its own right \cite{geonuanalyses}. 
As far as the precision measurement of 
the neutrino oscillation parameters is concerned, 
the events due to $\bar\nu^{geo}_e$ can be an 
important background 
and can lead to an error in the 
measured value of $\sss$.

  Since the $\bar\nu^{geo}_e$ have a maximum 
energy of $E_\anue = 3.26$ MeV
which corresponds to a prompt $e^+$ energy of 
only $E_{vis} = 2.48$ MeV, 
one way to avoid the uncertainty due to
$\bar\nu^{geo}_e$ is to implement
a prompt energy threshold of 2.6 MeV, 
as is done by the KamLAND collaboration.
In this paper we have followed the KamLAND approach. 
In this case the observed 
$e^+$ spectrum 
does not have any 
``contamination'' due to 
contributions from $\bar\nu^{geo}_e$. 
An alternative approach is to 
use the entire prompt $e^+$ energy spectrum 
in the analysis, taking the 
$\bar\nu^{geo}_e$ flux into account. 
Since the theoretical estimates 
on the $\bar\nu^{geo}_e$ flux 
are presently rather imprecise, 
one could let the $^{238}U$ and $^{232}Th$ 
$\bar\nu^{geo}_e$  flux normalisation 
vary as a free parameter. 
Both approaches have their merits 
and drawbacks. In the first approach 
(we use in this paper), 
while there are no additional 
uncertainties due to the unknown 
$\bar\nu^{geo}_e$ background, 
one has to contend with 
the experimental challenge of understanding and 
reducing the error associated with the 
prompt $e^+$ threshold energy.
{ In the case of KamLAND experiment
the uncertainty in the $e^+$ threshold energy
corresponds to a systematic error $\sim$2.3\%.}
In the second approach there 
is no prompt $e^+$ energy cut,
but one has to handle the uncertainty 
due to the lack of knowledge 
of $\bar\nu^{geo}_e$ background.
Keeping the $^{238}U-$ and $^{232}Th-$ 
$\bar\nu^{geo}_e$ flux normalisation 
as free parameter brings in extra 
error in the measurement on $\sss$.

   The key feature in the $\theta_{12}$ SPMIN reactor experiment
suggested in \cite{th12}, is the appearance of SPMIN in the 
observed $e^+$ spectrum. 
If the SPMIN appears at a prompt 
energy of $E_{vis} > 2.6$ MeV,  
implementing a threshold of $E_{vis} = 2.6$ MeV 
\footnote{This will increase the systematic 
uncertainty due to the error from the prompt 
$e^+$ energy cut. 
We have shown that the impact of the 
increase of the systematic uncertainty 
on the precision of $\sss$ determination
is relatively small as long as 
the systematic error does not exceed 5\%.}
and thus avoiding the $\bar\nu^{geo}_e$ background 
might permit to measure $\sss$ with the highest 
precision, achievable in the experiment under discussion.
If, on the other hand, SPMIN appears at $E_{vis} < 2.6$ MeV, 
the entire $e^+$ energy spectrum 
would have to be taken into 
account and in this case the  $\bar\nu^{geo}_e$ 
background cannot be avoided. 
For a given value of $\ms$, the position of 
the SPMIN in the  $\anue$ spectrum 
depends on the baseline of the experiment. 
For shorter baselines, SPMIN occurs at 
smaller energies. Therefore the
choice of the baseline of the 
experiment would determine 
whether one would have to take the 
$\bar\nu^{geo}_e$ background into account or not.

 The authors of \cite{minat12} have included the 
$\bar\nu^{geo}_e$ background in their analysis 
of the $\sss$ precision expected in the SADO
experiment in Japan 
with a baseline of $L = 54$ km. 
They conclude that for this experimental set-up, 
the  $\bar\nu^{geo}_e$ background does not have 
significant impact on the precision of $\sss$ measurement. 
For $L= 54$ km, the SPMIN is 
at $E_{vis} \cong 2.8$ MeV.
The uncertainty in the $\bar\nu^{geo}_e$ 
flux does not make much impact in this case
since $E_{vis} \cong 2.8$ MeV
is larger than the  background 
$\bar\nu^{geo}_e$ energies.
For shorter baselines the SPMIN will take place 
at $E_{vis} < 2.6$ MeV and the uncertainty due 
to the $\bar\nu^{geo}_e$ flux can
affect noticeably the precision of 
measurement of $\sss$.
\section{Conclusions}

   We have investigated the possibilities of high precision
measurement of the solar neutrino 
mixing angle $\theta_{12}$ in solar and reactor 
neutrino experiments. As a first step, we have analyzed the
improvements in the determination of $\sin^2\theta_{12}$,
which can be achieved with the expected increase
of statistics and reduction of systematic
errors in the currently operating 
solar and KamLAND experiments.
With the phase-III prospective data from SNO experiment 
included in the current global solar neutrino and 
KamLAND data, the uncertainty in the value of 
$\sin^2\theta_{12}$ is expected to diminish from 
24\% to 21\% at 3$\sigma$. If instead of 766.3 Ty, 
one uses simulated 3 kTy KamLAND data in the 
same analysis, the 3$\sigma$ error in 
$\sin^2\theta_{12}$ reduces to 18\%. 

 We next considered the potential of a generic 
LowNu $\nu-e$ elastic
scattering experiment, 
designed to measure the $pp$ solar 
neutrino flux, for high precision 
determination of $\sin^2\theta_{12}$. 
We examined the effect of including 
values of the $pp$ neutrino 
induced electron scattering rates in 
the $\chi^2$ analysis of the global solar neutrino data. 
Three representative values of the rates 
from the currently allowed 3$\sigma$ range were considered:
0.68, 0.72, 0.77. The error
in the measured rate was varied from 1\% to 5\%.
By adding the $pp$ flux data in the analysis,
the error in $\sin^2\theta_{12}$ determination
reduces to 14\% (19\%) at 3$\sigma$ for 1\% (3\%) uncertainty
in the measured $pp$ rate. Performing a similar 
three-neutrino oscillation analysis we found that, 
as a consequence of the uncertainty on $\sin^2\theta_{13}$,
the error on the value of $\sin^2\theta_{12}$ increases 
correspondingly to 17\% (21\%). 

  We also studied the possibility of a 
high precision determination of
$\sin^2\theta_{12}$ in a reactor experiment
with a baseline corresponding to a 
Survival Probability MINimum (SPMIN).
We showed that in a $L \sim 60 $ km experiment with statistics
of $\sim$60 GWkTy and systematic error 
of 2\%, $\sin^2\theta_{12}$  could be measured 
with an uncertainty of 2\% (6\%) at 1$\sigma$ (3$\sigma$). 
The inclusion of the $\sin^2\theta_{13}$ uncertainty
in the analysis changes this error to 3\%  (9\%).
An independent determination of $\ms$ 
with sufficiently high accuracy would allow,
as we have shown,
$\sss$ to be measured with the highest precision 
over a relatively wide range of baselines.
We investigated in detail the dependence of the
precision on $\sin^2\theta_{12}$ which can 
be achieved in such an experiment
on the baseline, statistics and systematic error. 
More specifically, with the increase of the statistics
from 20 GWkTy to 60 GWkTy, the error diminishes from
3\% (10\%) to 2\% (6\%) at 1$\sigma$ (3$\sigma$).
For statistics of (60 - 70) GWkTy, the increase of 
the systematic error from 2\% to 5\% leads to an increase
in the uncertainty in  $\sin^2\theta_{12}$
from 6\% to 9\% at 3$\sigma$.

   We have found that 
the effect of $\sin^2\theta_{13}$ uncertainty
on the $\sin^2\theta_{12}$ determination in LowNu $pp$ and 
SPMIN reactor experiments considered is 
considerably smaller than naively expected.

   The results of our analyses for the currently 
running, the proposed LowNu and future reactor experiments
show that the most precise determination of $\sin^2\theta_{12}$
can be achieved in a dedicated reactor experiment 
with a baseline tuned to SPMIN associated with
$\ms \equiv \Delta m^2_{\odot}$. 

\vskip 0.4cm
{\bf Acknowledgments:} We would like to thank Y. Suzuki, 
A. Suzuki, M. Nakahata, F. Suekane,
and C. Pe\~na-Garay for useful discussions.
This work was supported by the Italian INFN 
under the program ``Fisica Astroparticellare'' 
(S.C. and S.T.P.).

\vspace{-0.5cm}


\end{document}